\documentclass[manyauthors,nocleardouble,COMPASS]{cernphprep}
\usepackage{graphicx}
\usepackage{subfig}
\usepackage[percent]{overpic}
\usepackage[numbers, square, comma, sort&compress]{natbib}

\usepackage[english]{babel}

\usepackage{graphicx}
\usepackage{rotating}
\usepackage{multirow}
\usepackage{amsmath}
\usepackage{amssymb}
\usepackage{array}
\usepackage{times}
\usepackage{colordvi}

\usepackage{hyperref}

\def\lapproxeq{\lower .7ex\hbox{$\;\stackrel{\textstyle
<}{\sim}\;$}}
\def\gapproxeq{\lower .7ex\hbox{$\;\stackrel{\textstyle
>}{\sim}\;$}}

\begin{document}
\selectlanguage{english}

\begin{titlepage}
\PHnumber{2013--052}
\PHdate{March 25, 2013}

\title{Study of $\Sigma$(1385) and $\Xi$(1321) hyperon and antihyperon production in deep inelastic muon scattering}
 
\Collaboration{The COMPASS Collaboration}
\ShortAuthor{The COMPASS Collaboration}

\begin{abstract}
Large samples of $\Lambda$, $\Sigma(1385)$ and $\Xi(1321)$ hyperons produced in the
deep-inelastic muon scattering off a $^6$LiD target were collected with the
COMPASS experimental setup at CERN. The relative yields of $\Sigma(1385)^+$, 
$\Sigma(1385)^-$, $\bar{\Sigma}(1385)^-$, $\bar{\Sigma}(1385)^+$, $\Xi(1321)^-$, 
and $\bar{\Xi}(1321)^+$ hyperons decaying into $\Lambda$($\bar{\Lambda}$)$\pi$
were measured. The ratios of heavy-hyperon to $\Lambda$ and heavy-antihyperon to $\bar{\Lambda}$ were 
found to be in the range 3.8\% to 5.6\% with a
relative uncertainty of about 10\%. They were used to tune the parameters
relevant for strange particle production of the LEPTO Monte Carlo generator.
\end{abstract}

\vfill
\Submitted{(Submitted to The European Physical Journal C)}
\end{titlepage}

{\pagestyle{empty}
%
%

\section*{The COMPASS Collaboration}
\label{app:collab}
\renewcommand\labelenumi{\textsuperscript{\theenumi}~}
\renewcommand\theenumi{\arabic{enumi}}
\begin{flushleft}
C.~Adolph\Irefn{erlangen},
M.G.~Alekseev\Irefn{triest_i},
V.Yu.~Alexakhin\Irefn{dubna},
Yu.~Alexandrov\Irefn{moscowlpi}\Deceased,
G.D.~Alexeev\Irefn{dubna},
A.~Amoroso\Irefn{turin_u},
A.~Austregesilo\Irefnn{cern}{munichtu},
B.~Bade{\l}ek\Irefn{warsawu},
F.~Balestra\Irefn{turin_u},
J.~Barth\Irefn{bonnpi}
% ...
\end{flushleft}

%
%

\begin{Authlist}
\item \Idef{bielefeld}{Universit\"at Bielefeld, Fakult\"at f\"ur Physik, 33501 Bielefeld, Germany\Arefs{f}}
\item \Idef{bochum}{Universit\"at Bochum, Institut f\"ur Experimentalphysik, 44780 Bochum, Germany\Arefs{f}}
\item \Idef{bonniskp}{Universit\"at Bonn, Helmholtz-Institut f\"ur  Strahlen- und Kernphysik, 53115 Bonn, Germany\Arefs{f}}
\item \Idef{bonnpi}{Universit\"at Bonn, Physikalisches Institut, 53115 Bonn, Germany\Arefs{f}}
\item \Idef{brno}{Institute of Scientific Instruments, AS CR, 61264 Brno, Czech Republic\Arefs{g}}
\item \Idef{calcutta}{Matrivani Institute of Experimental Research \& Education, Calcutta-700 030, India\Arefs{h}}
\item \Idef{dubna}{Joint Institute for Nuclear Research, 141980 Dubna, Moscow region, Russia\Arefs{i}}
\item \Idef{erlangen}{Universit\"at Erlangen--N\"urnberg, Physikalisches Institut, 91054 Erlangen, Germany\Arefs{f}}
\item \Idef{freiburg}{Universit\"at Freiburg, Physikalisches Institut, 79104 Freiburg, Germany\Arefs{f}}
\item \Idef{cern}{CERN, 1211 Geneva 23, Switzerland}
\item \Idef{liberec}{Technical University in Liberec, 46117 Liberec, Czech Republic\Arefs{g}}
\item \Idef{lisbon}{LIP, 1000-149 Lisbon, Portugal\Arefs{j}}
\item \Idef{mainz}{Universit\"at Mainz, Institut f\"ur Kernphysik, 55099 Mainz, Germany\Arefs{f}}
\item \Idef{miyazaki}{University of Miyazaki, Miyazaki 889-2192, Japan\Arefs{k}}
\item \Idef{moscowlpi}{Lebedev Physical Institute, 119991 Moscow, Russia}
\item \Idef{munichlmu}{Ludwig-Maximilians-Universit\"at M\"unchen, Department f\"ur Physik, 80799 Munich, Germany\Arefs{f}\Arefs{l}}
\item \Idef{munichtu}{Technische Universit\"at M\"unchen, Physik Department, 85748 Garching, Germany\Arefs{f}\Arefs{l}}
\item \Idef{nagoya}{Nagoya University, 464 Nagoya, Japan\Arefs{k}}
\item \Idef{praguecu}{Charles University in Prague, Faculty of Mathematics and Physics, 18000 Prague, Czech Republic\Arefs{g}}
\item \Idef{praguectu}{Czech Technical University in Prague, 16636 Prague, Czech Republic\Arefs{g}}
\item \Idef{protvino}{State Research Center of the Russian Federation, Institute for High Energy Physics, 142281 Protvino, Russia}
\item \Idef{saclay}{CEA IRFU/SPhN Saclay, 91191 Gif-sur-Yvette, France}
\item \Idef{telaviv}{Tel Aviv University, School of Physics and Astronomy, 69978 Tel Aviv, Israel\Arefs{m}}
\item \Idef{triest_i}{Trieste Section of INFN, 34127 Trieste, Italy}
\item \Idef{triest}{University of Trieste, Department of Physics and Trieste Section of INFN, 34127 Trieste, Italy}
\item \Idef{triestictp}{Abdus Salam ICTP and Trieste Section of INFN, 34127 Trieste, Italy}
\item \Idef{turin_u}{University of Turin, Department of Physics and Torino Section of INFN, 10125 Turin, Italy}
\item \Idef{turin_i}{Torino Section of INFN, 10125 Turin, Italy}
\item \Idef{turin_p}{University of Eastern Piedmont, 15100 Alessandria,  and Torino Section of INFN, 10125 Turin, Italy}
\item \Idef{warsaw}{National Centre for Nuclear Research, 00-681 Warsaw, Poland\Arefs{n} }
\item \Idef{warsawu}{University of Warsaw, Faculty of Physics, 00-681 Warsaw, Poland\Arefs{n} }
\item \Idef{warsawtu}{Warsaw University of Technology, Institute of Radioelectronics, 00-665 Warsaw, Poland\Arefs{n} }
\item \Idef{yamagata}{Yamagata University, Yamagata, 992-8510 Japan\Arefs{k} }
\end{Authlist}
%
%
\vspace*{-\baselineskip}\renewcommand\theenumi{\alph{enumi}}
\begin{Authlist}
\item \Adef{a}{Also at IST, Universidade T\'ecnica de Lisboa, Lisbon, Portugal}
\item \Adef{bb}{Supported by the DFG Research Training Group Programme 1102  ``Physics at Hadron Accelerators''}
\item \Adef{b}{Also at Chubu University, Kasugai, Aichi, 487-8501 Japan\Arefs{k}}
\item \Adef{c}{Also at KEK, 1-1 Oho, Tsukuba, Ibaraki, 305-0801 Japan}
\item \Adef{cc}{Also at Moscow Institute of Physics and Technology, Moscow Region, 141700, Russia}
\item \Adef{y}{present address: National Science Foundation, 4201 Wilson Boulevard, Arlington, VA 22230, United States}
\item \Adef{x}{present address: RWTH Aachen University, III. Physikalisches Institut, 52056 Aachen, Germany}
\item \Adef{e}{Also at GSI mbH, Planckstr.\ 1, D-64291 Darmstadt, Germany}
\item \Adef{f}{Supported by the German Bundesministerium f\"ur Bildung und Forschung}
\item \Adef{g}{Supported by Czech Republic MEYS Grants ME492 and LA242}
\item \Adef{h}{Supported by SAIL (CSR), Govt.\ of India}
\item \Adef{i}{Supported by CERN-RFBR Grants 08-02-91009}
\item \Adef{j}{\raggedright Supported by the Portuguese FCT - Funda\c{c}\~{a}o para a Ci\^{e}ncia e Tecnologia, COMPETE and QREN, Grants CERN/FP/109323/2009, CERN/FP/116376/2010 and CERN/FP/123600/2011}
\item \Adef{k}{Supported by the MEXT and the JSPS under the Grants No.18002006, No.20540299 and No.18540281; Daiko Foundation and Yamada Foundation}
\item \Adef{l}{Supported by the DFG cluster of excellence `Origin and Structure of the Universe' (www.universe-cluster.de)}
\item \Adef{m}{Supported by the Israel Science Foundation, founded by the Israel Academy of Sciences and Humanities}
\item \Adef{n}{Supported by the Polish NCN Grant DEC-2011/01/M/ST2/02350}
\item [{\makebox[2mm][l]{\textsuperscript{*}}}] Deceased
\end{Authlist}

\clearpage
}

\maketitle

\section {Introduction}
\label{intro}
The study of hyperon production in deep inelastic scattering (DIS) is important
for a better understanding of the role of strange quarks in the nucleon
structure and in the hadronization process. The lightest hyperon, the $\Lambda$
baryon, was studied in most detail. In addition to $\Lambda$ from direct
production, a significant fraction of $\Lambda$ particles originates from the
decay of heavier hyperons such as $\Sigma^0$, $\Sigma^*$, or $\Xi$. 
The notation for $\Sigma(1385)$ and $\Xi(1321)$ will be used without indicating mass
values, and with the "*" symbol for $\Sigma(1385)$ in order to distinguish the
$J^{P}=3/2^{+}$ $\Sigma$ hyperons from the $J^{P}=1/2^{+}$ ones. 
An indirect $\Lambda$ production from hyperons decays is 
also included in the measurements of the longitudinal spin transfer to the $\Lambda$
hyperon in polarised DIS~\cite{HERMES-lambda,Compass:09}. Using a Monte Carlo
simulation based on the Lund string fragmentation model~\cite{LEPTO}, the
authors of Ref.~\cite{HERMES-lambda} have estimated that only about 40\% of the
produced $\Lambda$ baryons originate from direct string fragmentation.

The production of $\Sigma^0$, $\Sigma^{*+}$, $\Sigma^{*-}$ and $\Xi^-$ hyperons
with neutrino beams was reported by the NOMAD Collaboration
~\cite{NOMAD-resonances1}. Information on the production of these heavy
hyperons with muon or electron beams is still missing. To our knowledge, the
production of the antiparticles, $\bar \Sigma^{*-}$, $\bar{\Sigma}^{*+}$, and
$\bar\Xi$ has never been studied in DIS. New data are hence required in order
to produce reliable numerical estimate of heavy hyperon production rates in DIS.

In this Paper, the production rates of $\Sigma^{*+}$, $\Sigma^{*-}$, $\Xi$ and
their antiparticles are presented and compared to those of $\Lambda$ and
$\bar{\Lambda}$ hyperons. The resulting values are used to constrain the
parameters of the JETSET package whith is embedded in the LEPTO Monte Carlo
generator.

\section{The experimental setup}
 \label{Sec_exper}
The data used in the present analysis were collected by the COMPASS
Collaboration at CERN during the years 2003--2004. The experiment was performed
at the CERN M2 muon beam line. The $\mu^{+}$ beam intensity was $2\cdot10^8$
per spill of 4.8~s, with a cycle time of 16.8~s. The average beam momentum was
160~GeV/{\it c}. The $\mu^{+}$ beam is naturally polarised by the weak decays
of the parent hadrons.

The beam traverses two cylindrical cells of a polarised $^{6}$LiD target, both
of 60 cm length and 3 cm diameter. The target material in the neighbouring
cells is polarised longitudinally in opposite directions with respect to the
beam. However, the target polarisation values are not used in this study. The
data from both target cells and polarisations are combined.

The COMPASS experimental setup was designed to detect both scattered muons and
produced hadrons in wide momentum and angular ranges. It consists of two
spectrometer stages, each comprising a large-aperture dipole magnet. The
aperture of the target magnet limits the acceptance to $\pm 70$~mrad at the
upstream end of the target. Muons are identified in large area tracking
detectors and scintillators downstream of concrete or iron muon filters.
Hadrons are detected in two hadron calorimeters installed upstream of the muon
filters.

Data recording is activated by inclusive and semi-inclusive triggers indicating
the presence of a scattered muon and emitted hadrons, respectively. The trigger
system covers a wide range of $Q^2$ from quasi-real photoproduction to deep
inelastic interactions. A more detailed description of the COMPASS apparatus
can be found in Ref.~\cite{setup}.

\section{$\Lambda$ and $\bar{\Lambda}$ hyperon samples }
 \label{Sec_sample}
The event selection requires a reconstructed interaction vertex that is defined
by the incoming and the scattered muon and located within the target. DIS
events are selected by cuts on the four-momentum squared of the virtual photon,
$Q^{2}>1~$~(GeV/{\it c})$^{2}$, and on the fractional energy $y$ of the virtual
photon, $0.2<y<0.9$. The latter cut removes events with large radiative
corrections at large $y$ and with poorly reconstructed kinematic variables at
low $y$. The resulting data sample consists of $3.12\cdot10^8$ events. The
$\Lambda$ and $\bar{\Lambda}$ hyperons are identified by their decays into
$p\pi^-$ and $\bar{p}\pi^+$, respectively. In order to evaluate possible
systematic effects, decays of $K^0_s$ into $\pi^- \pi^+$ were also analysed. 
The $K^0_s$ background was already disscused in details in Ref.~\cite{Compass:09}.
Candidate events for $\Lambda$, $\bar{\Lambda}$ and $K^0_s$ were selected by
requiring that two hadron tracks form a secondary vertex located within a 105~cm
long fiducial region starting 5~cm downstream of the target. Outside this
region, decay-hadron tracks cannot be reconstructed with sufficient resolution.
Vertices with identified muons or electrons were removed. Only hadrons with
momenta larger than 1~GeV/{\it c} were retained, guaranteeing a good
reconstruction efficiency. A further cut was imposed on the transverse momentum
$p_T$ of the decay products with respect to the hyperon direction,
$p_T>23$~MeV/{\it c}, in order to reject $e^{+}e^{-}$ pairs from
$\gamma$-conversion. Using the Feynman variable $x_F$, the
$\Lambda$($\bar{\Lambda}$) candidates were selected in the current fragmentation
region requiring $x_F>0.05$.

\begin{figure}
\begin{center}
\includegraphics[width=0.49\textwidth]{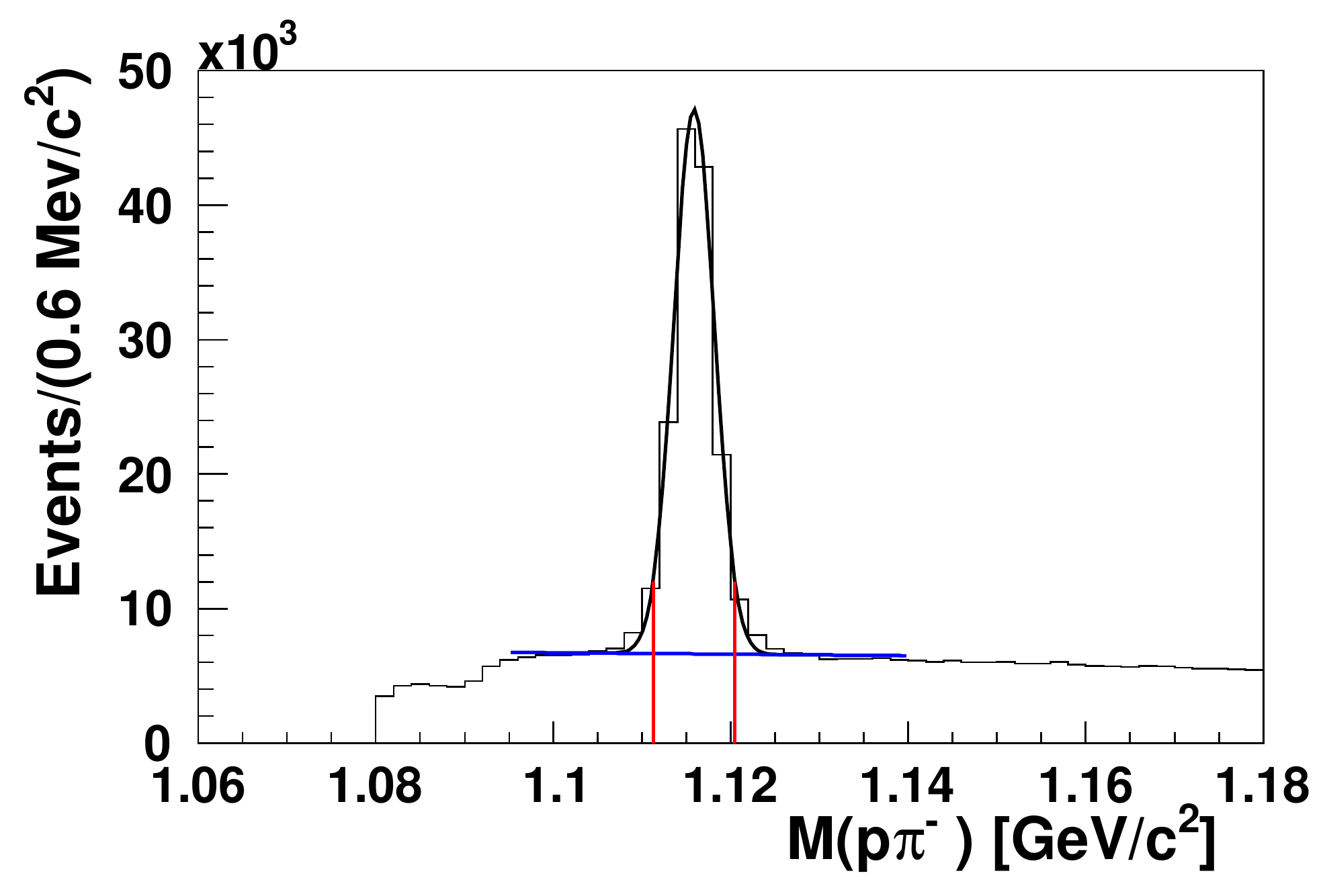} \hfill
\includegraphics[width=0.49\textwidth]{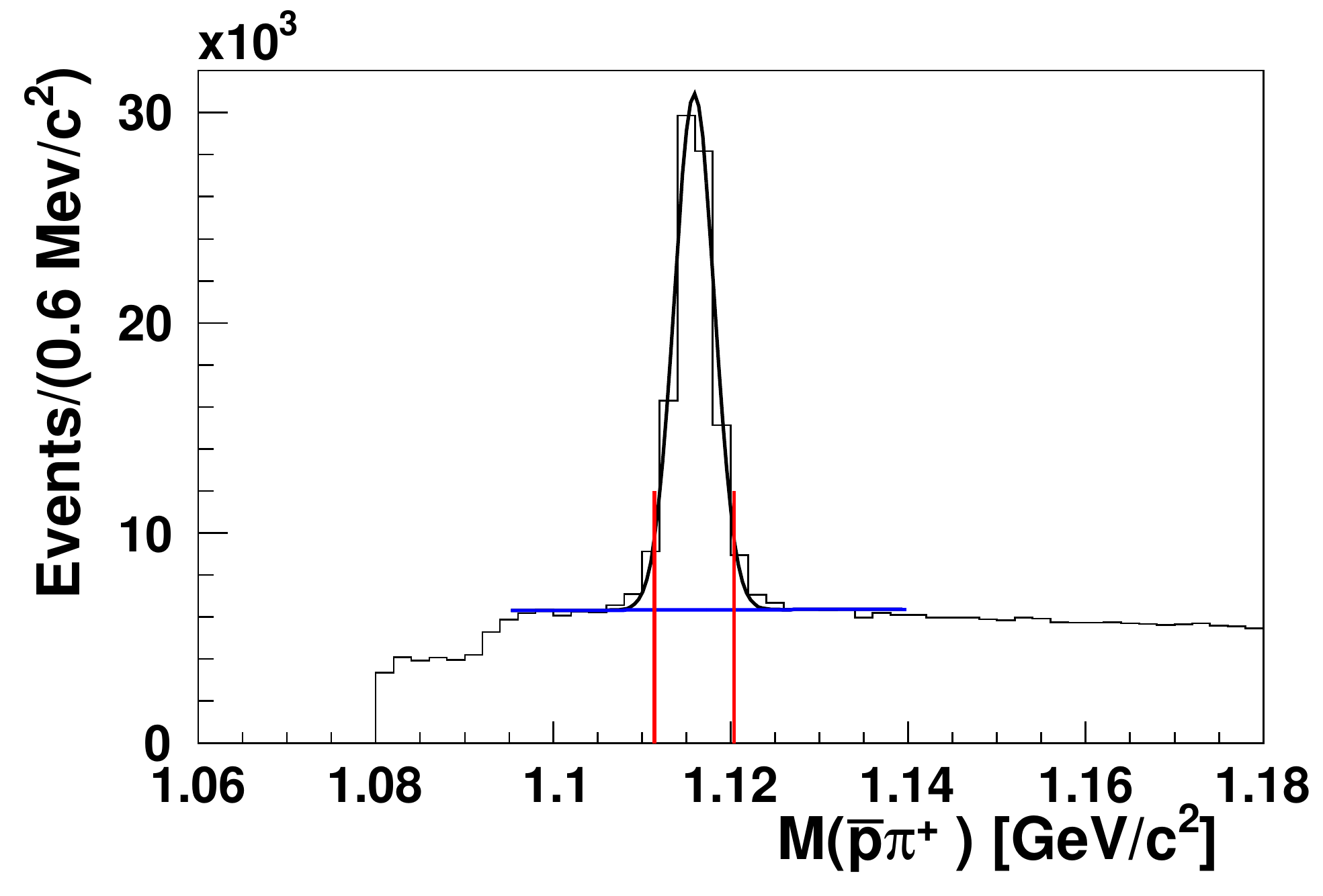}
\caption{The $p\pi^-$ (top) and $\bar{p}\pi^+$ (bottom) invariant mass
distributions. The total numbers of $\Lambda$ and $\bar{\Lambda}$ determined
within the fit interval (see text) are $N(\Lambda)=112449 \pm 418$ and 
$N(\bar \Lambda)=66685 \pm 350$. The vertical lines mark the $\pm 2 \sigma$
intervals of the $\Lambda$($\bar{\Lambda}$)~signals used for the $\Sigma^{*}$,
$\Xi$ and their antiparticle search. }
\label{inv-mass-nocoll}
\end{center}
\end{figure}

The invariant mass distributions for two-hadron events, which
are assumed to be either $p\pi^-$ or $\bar{p}\pi^+$ pairs, are shown in
Fig.~\ref{inv-mass-nocoll}. The distributions were fitted in the interval
$1.095 \div 1.140$~GeV/$c^2$ with a sum of a Gaussian function for the signal and a
third-order polynomial for the background. The total numbers of $\Lambda$ and
$\bar{\Lambda}$ hyperons, represents an improvement of an
order of magnitude\footnote{The samples used in the $\Lambda$
analysis~\cite{Compass:09} and in this Paper are different due to different
cuts.} with respect to previous experiments~\cite{HERMES-lambda,
NOMAD-resonances2, NOMAD-resonances3, E665, STAR}. The invariant mass
resolutions for $\Lambda$ and $\bar \Lambda$ are $2.22\pm0.01$~MeV/$c^2$ and
$2.21\pm0.01$~MeV/$c^2$, respectively.

\begin{figure}
\begin{center}
\includegraphics[width=0.49\textwidth]{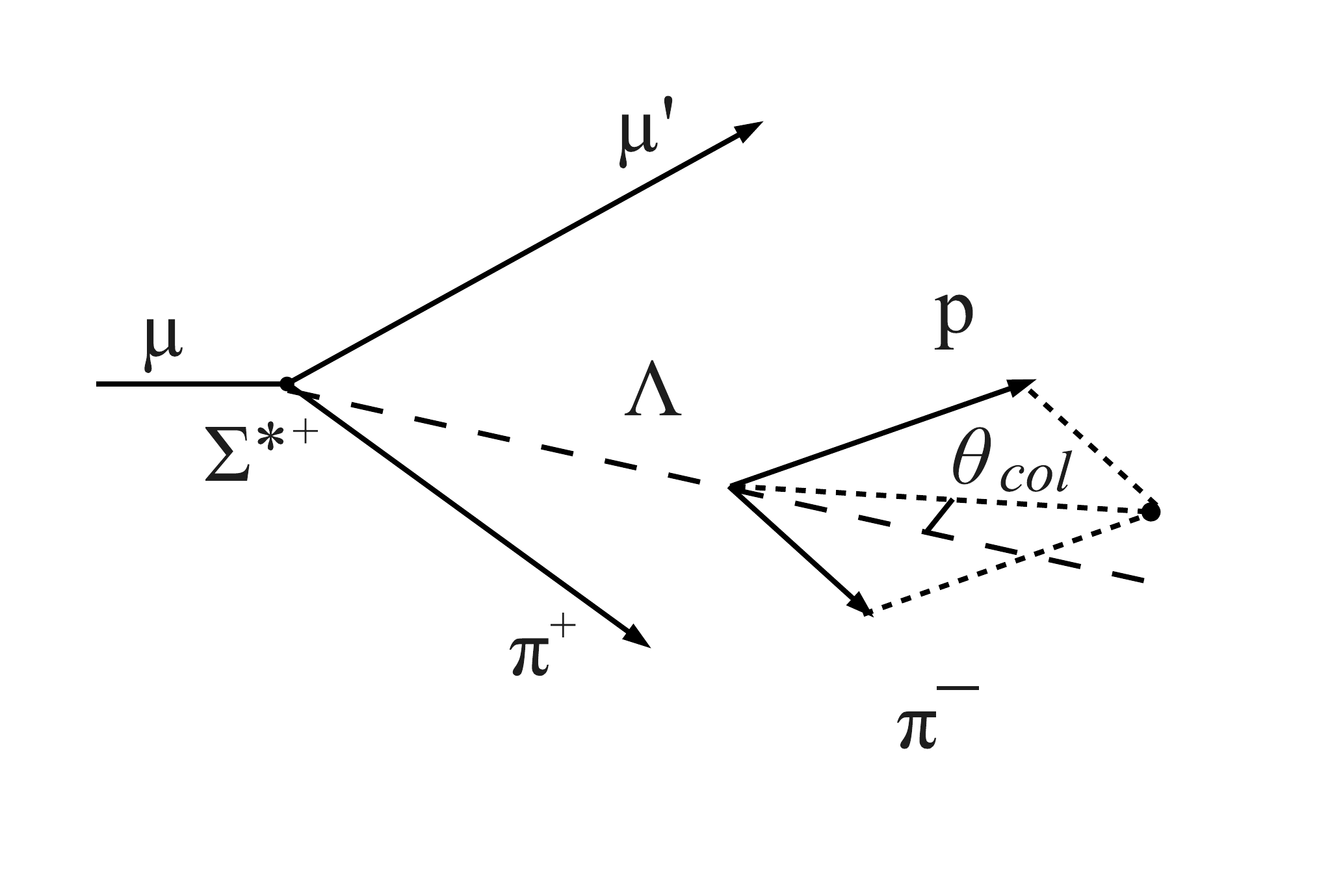} \hfill
\includegraphics[width=0.49\textwidth]{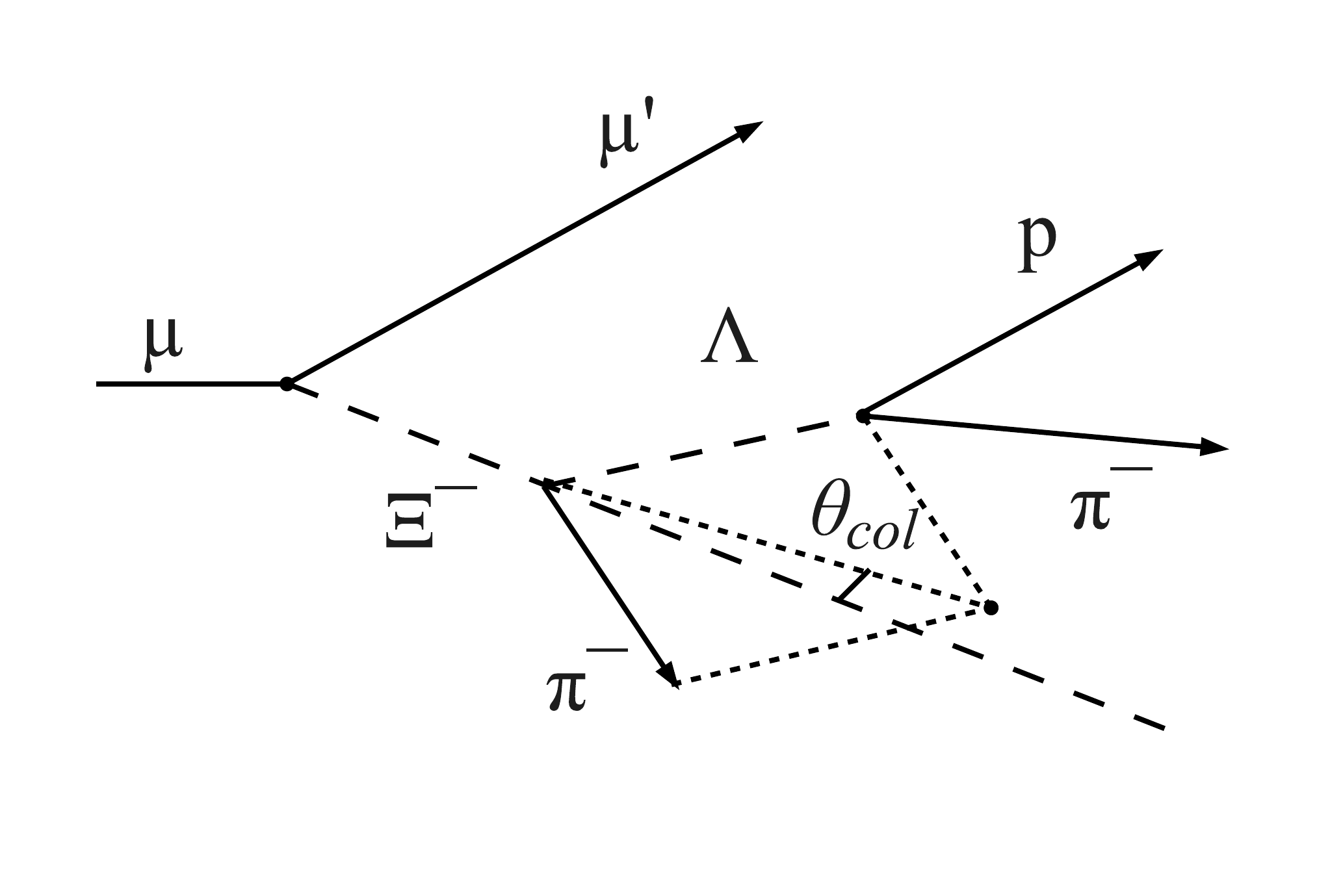} 
\caption{Schematic picture of the $\Sigma^{*+}$ strong decay (top) and of the
  $\Xi^{-}$ weak decay (bottom).}
\label{scheme}
\end{center}
\end{figure}

The $\Lambda$ hyperons in the resulting event samples are either directly
produced or originate from the decay of heavier hyperons. The $\Sigma^{*}$ and
$\Xi$ hyperons and their antiparticle partners decay with fractions of 87.5\% or
99.9\%, respectively, into $\Lambda$($\bar{\Lambda}$)$\pi^\pm$. The production
and the decay of the $\Sigma^{*}$ hyperon is illustrated in Fig.~\ref{scheme} (top).
Since the $\Sigma^{*}$ decays via strong interaction, the production and decay
vertices are indistinguishable. The secondary vertex is the signature of the
$\Lambda$($\bar{\Lambda}$) weak decay. In contrast, the $\Xi$ hyperon decays
via weak interaction (Fig.~\ref{scheme} (bottom)), such that the decay vertex is clearly
separated from the production vertex.

\section{\protect$\Sigma^{*}$ and \protect$\Xi$ hyperon samples}
\label{sec:results}

\begin{figure*}
\begin{center}
\begin{tabular}{cc}
\renewcommand{\arraystretch}{0.1}
\includegraphics[width=0.47\textwidth]{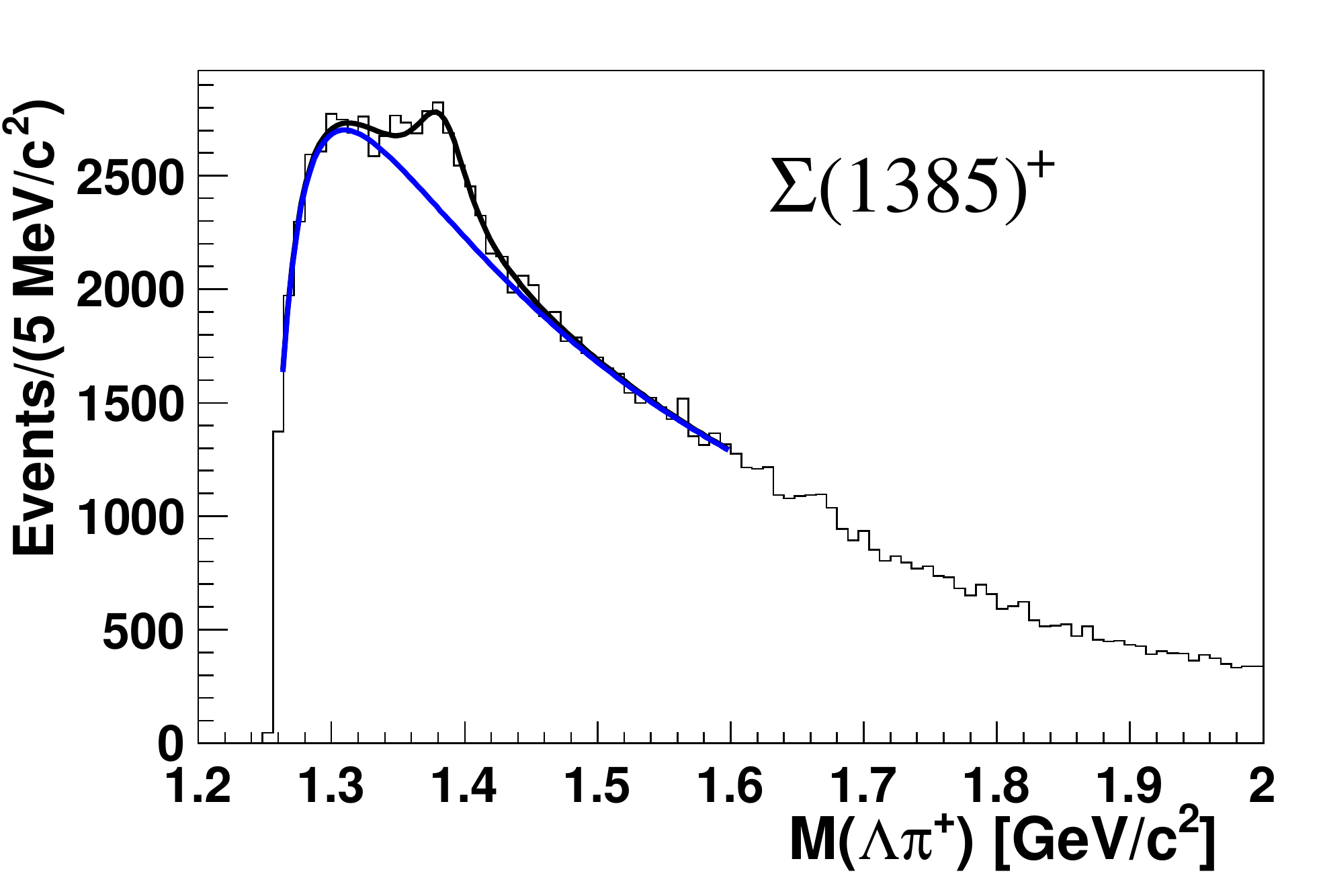} &
\includegraphics[width=0.47\textwidth]{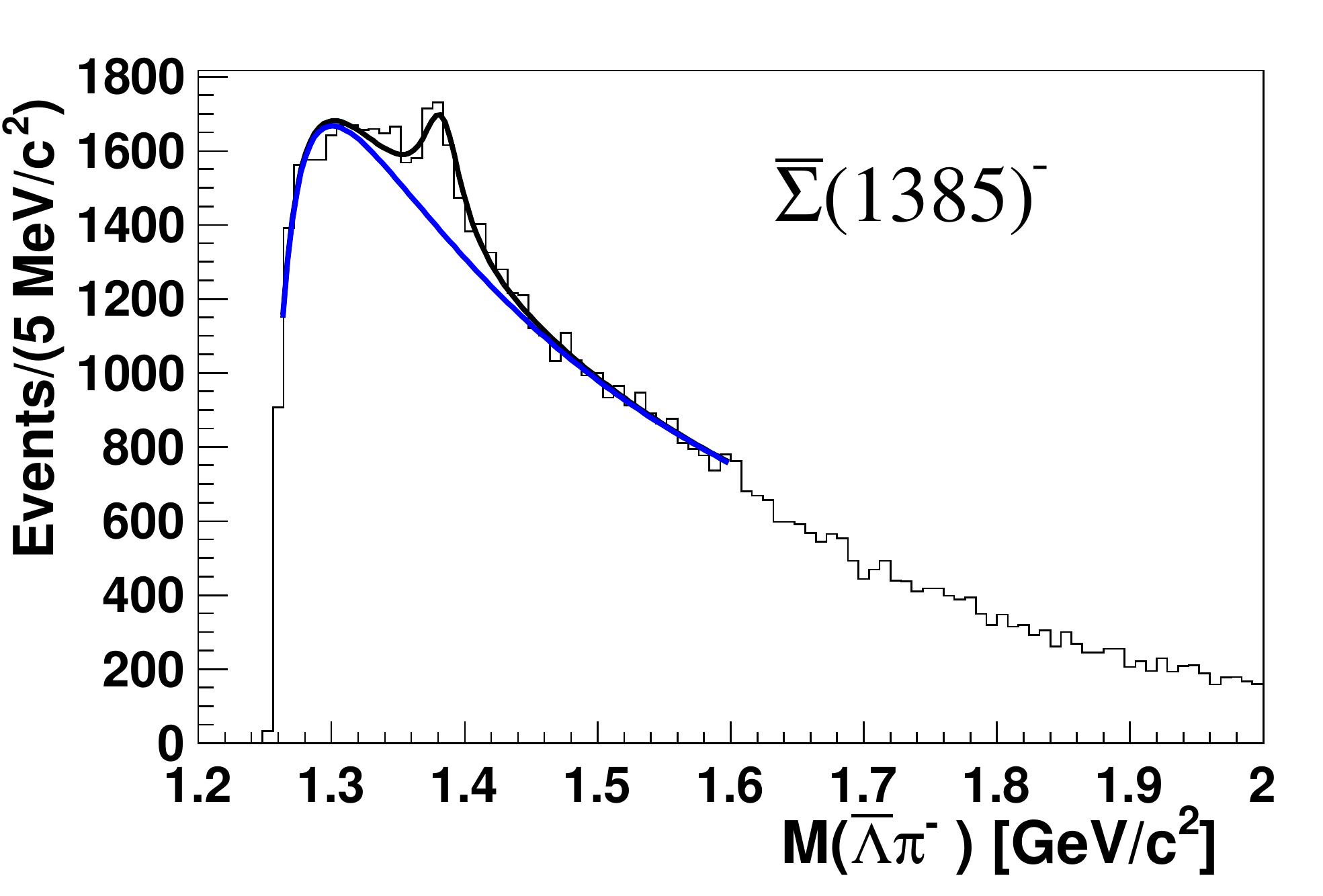} \\
a) & b) \\
\includegraphics[width=0.47\textwidth]{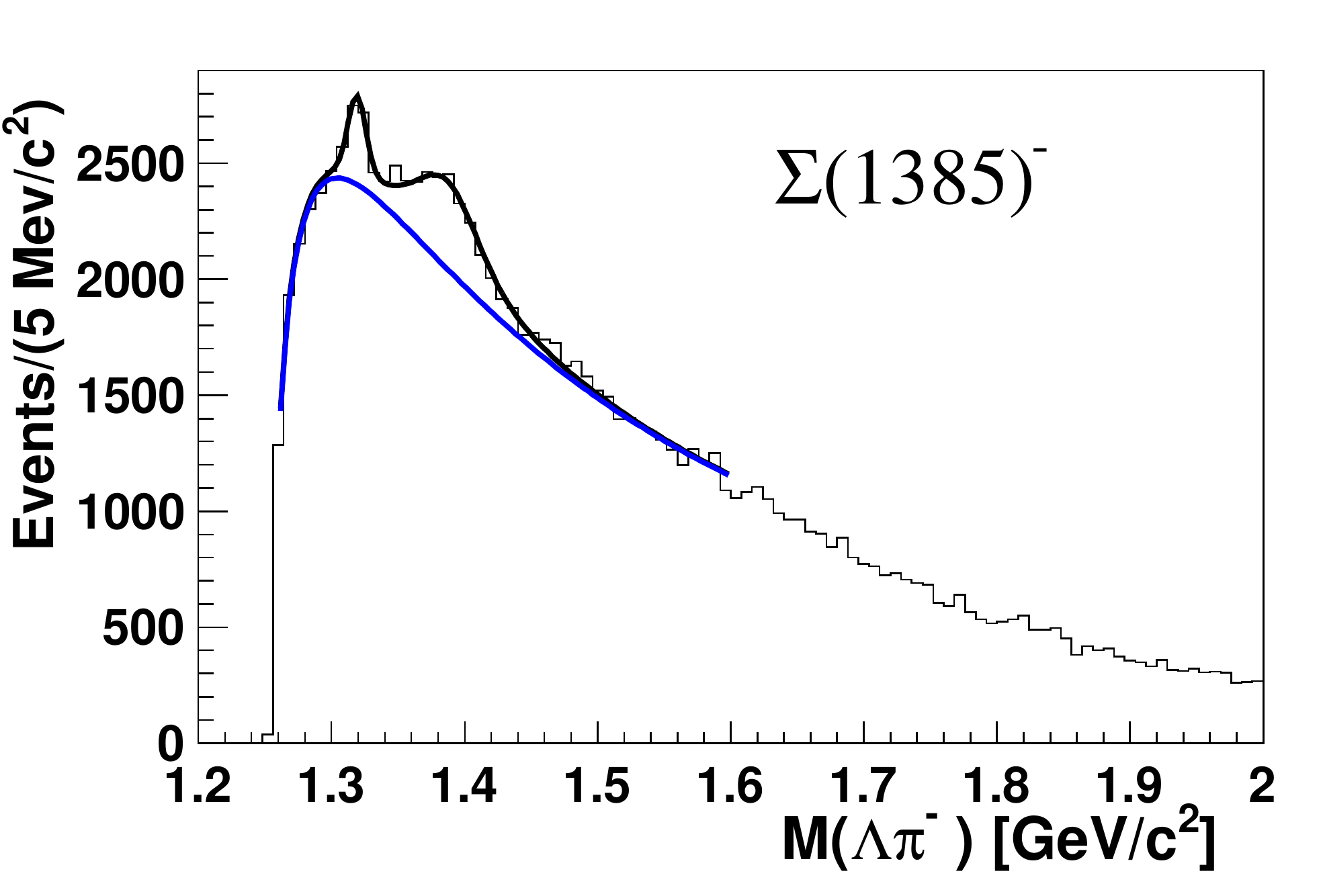} &
\includegraphics[width=0.47\textwidth]{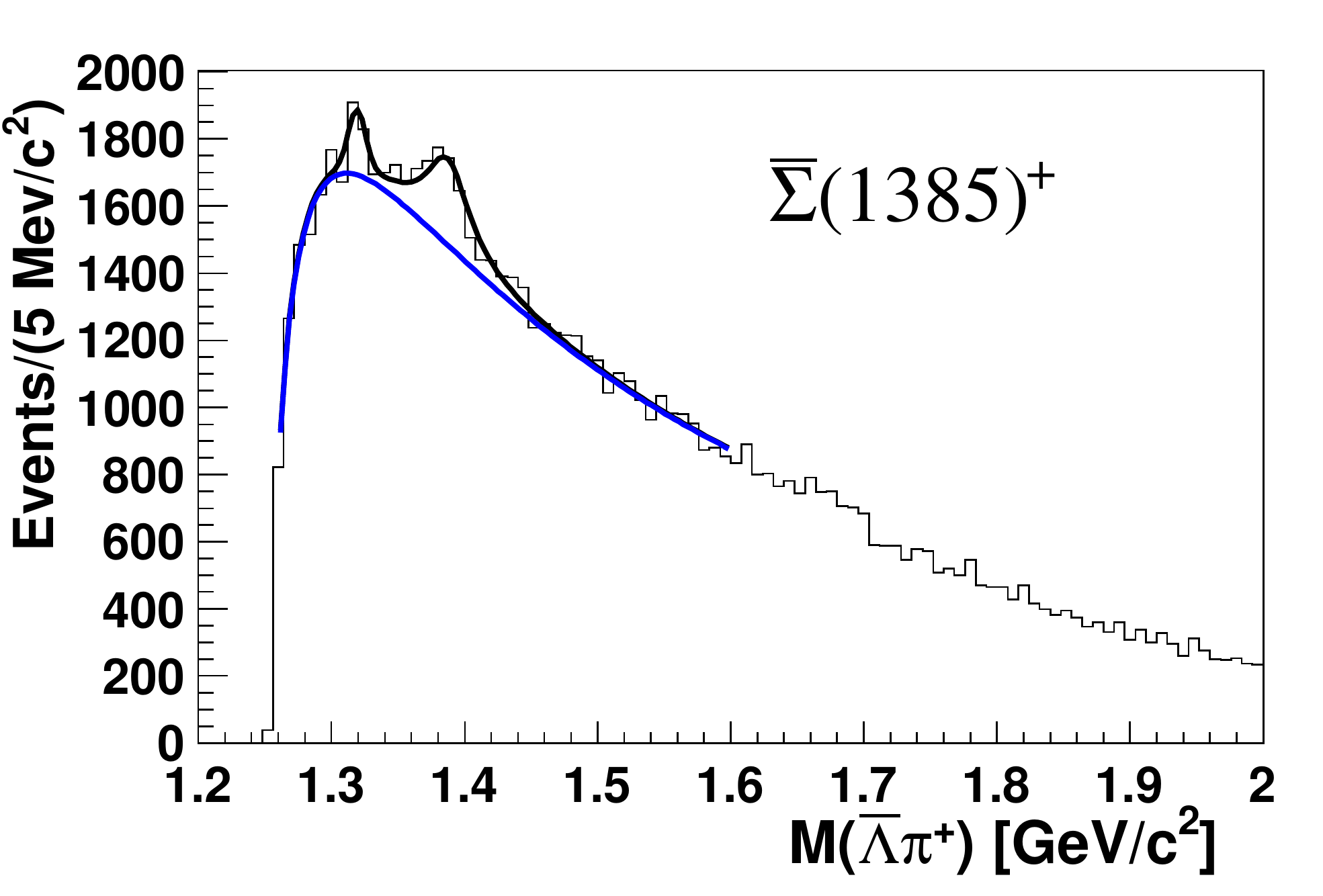} \\
c) & d) \\
\end{tabular}
\caption{The $\Lambda\pi$ invariant mass distributions. The solid lines
represent the signal plus backgroud and the background only obtained from the
fit. The signals include peaks for the following candidates: a)
$\Sigma^{*+}\rightarrow\Lambda\pi^{+}$; b) $\bar
\Sigma^{*-}\rightarrow\bar{\Lambda}\pi^{-}$; c)
$\Sigma^{*-}\rightarrow\Lambda\pi^{-}$ and $\Xi^{-}\rightarrow\Lambda\pi^{-}$;
d) $\bar \Sigma^{*+}\rightarrow\bar{\Lambda}\pi^{+}$ and $\bar
\Xi^{+}\rightarrow\bar{\Lambda}\pi^{+}$. The number of $\Sigma^{*}$ resulting
from the fits are: $N(\Sigma^{*+})=3631 \pm 333$, $N(\Sigma^{*-})=2970 \pm 490$,
$N(\bar{\Sigma}^{*-})=2173 \pm 222$ and $N(\bar{\Sigma}^{*+})=1889\pm 265$.}
\label{sigma_mass}
\end{center}
\end{figure*}

The search for $\Sigma^{*}$ hyperons was performed using the samples obtained
after a collinearity cut. This cut requires that the angle $\theta_{col}$
between the $\Lambda$ momentum and the line connecting the primary and the
secondary vertex is smaller than $0.01$ rad. It ensures predominant selection
of $\Lambda$ baryons pointing to the primary vertex and removes only 10$\%$ of
their total yield. The $p\pi^-$($\bar p\pi^+$) pairs within a $\pm 2\sigma$
mass interval from the mean value of the $\Lambda$($\bar{\Lambda}$)~peak were
then combined with a charged track from the primary vertex, which is assumed to
be a pion. All possible combinations were taken into account. The resulting
$\Lambda \pi$ invariant mass distributions are shown in Fig.~\ref{sigma_mass}.
The peaks for $\Sigma^{*+}$, $\bar{\Sigma}^{*-}$, $\Sigma^{*-}$, and
$\bar{\Sigma}^{*+}$ production are clearly visible. In the two bottom panels,
the small additional peaks of $\Xi^{-}$ and $\bar \Xi^{+}$ are also visible,
despite the fact that the $\Lambda$($\bar{\Lambda}$) from the decays of these
hyperons originate not from the primary vertex. The 1$\sigma$ mass resolutions
for $\Sigma^{*}$ and $\bar{\Sigma}^{*}$ agree within uncertainties:
$9.3\pm3.6$~MeV/$c^2$ for $\Sigma^{*+}$, $6.1\pm2.7$~MeV/$c^2$ for
$\bar{\Sigma}^{*-}$, $8.7\pm3.5$~MeV/$c^2$ for $\Sigma^{*-}$ and
$7.1\pm2.1$~MeV/$c^2$ for $\bar{\Sigma}^{*+}$.

\begin{figure}
\begin{center}
\includegraphics[width=0.49\textwidth]{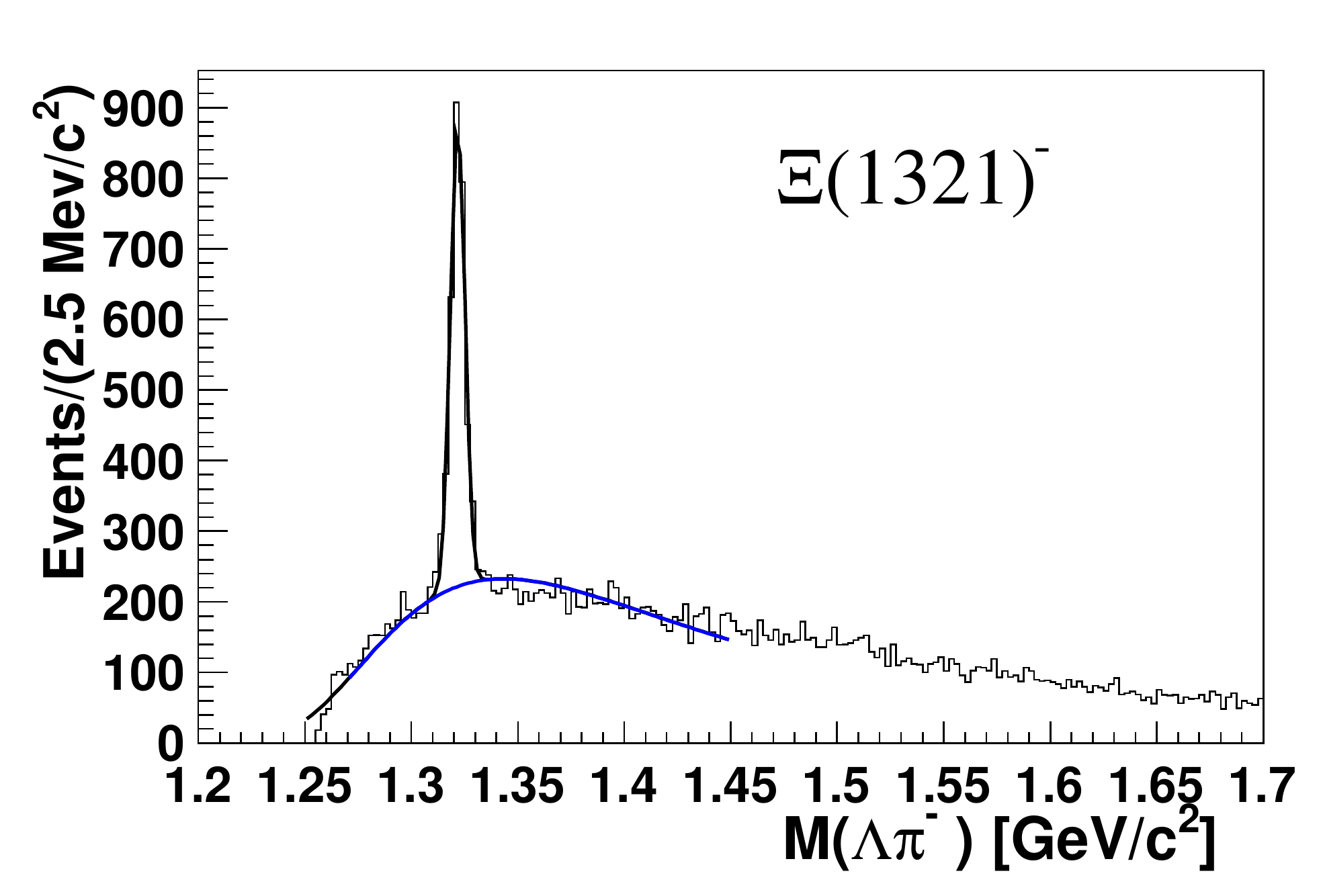} \hfill
\includegraphics[width=0.49\textwidth]{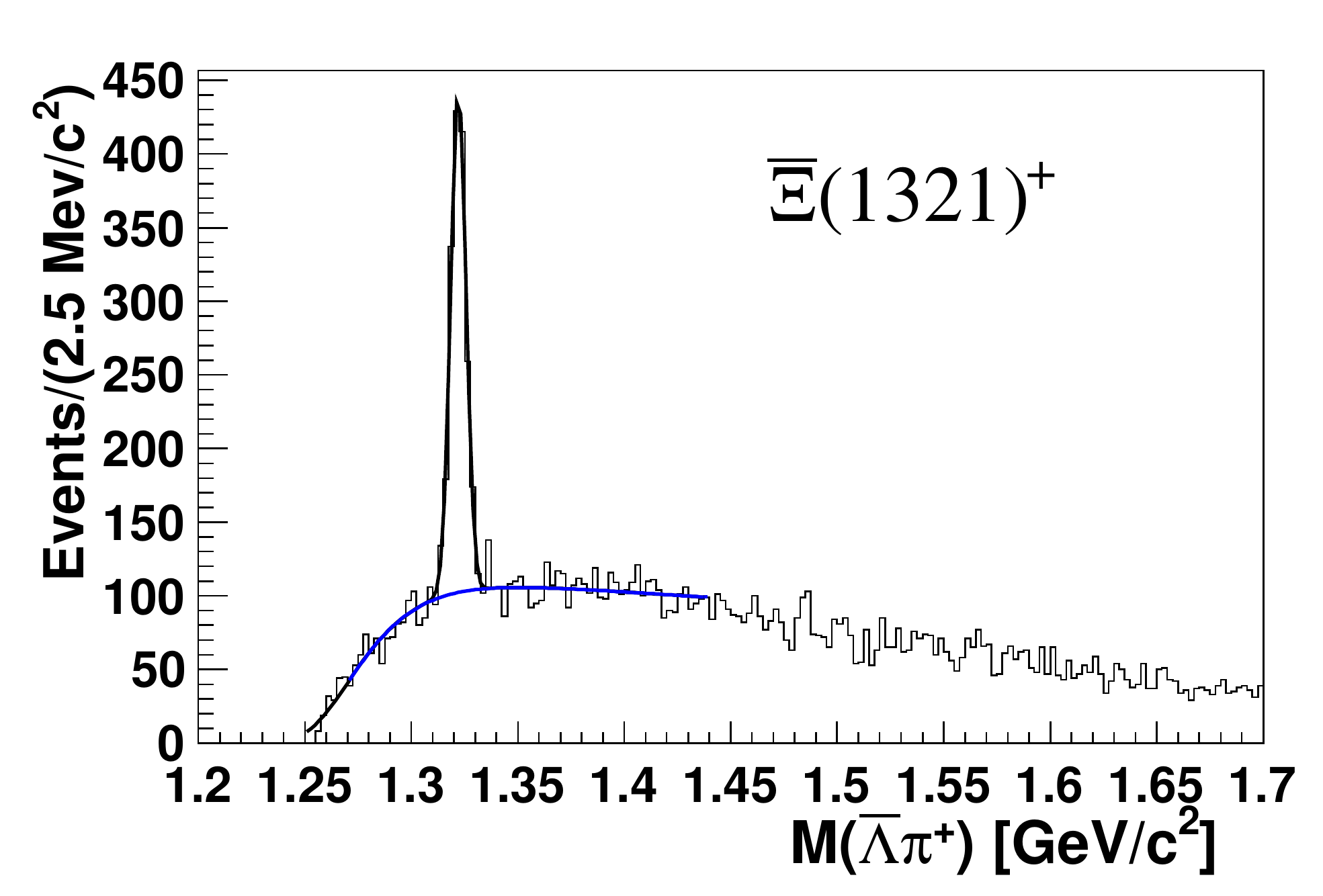} 
\caption{Invariant mass distributions for $\Lambda\pi^{-}$ (top) and
$\bar{\Lambda}\pi^{+}$ (bottom) pairs . The solid lines represent the signal plus
backgroud and the background only obtained from the fit. The peaks correspond
to $\Xi^{-}\rightarrow\Lambda\pi^{-}$ (top) and  $\bar
\Xi^{+}\rightarrow\bar{\Lambda}\pi^{+}$ (bottom) candidates. The estimated numbers of 
$\Xi$ hyperons are: $N(\Xi^{-}) = 2320 \pm 68$ and $N(\bar{\Xi}^{+}) = 1147 \pm
49$.}
\label{xi_mass_minus_before}
\end{center}
\end{figure}

An alternative method was chosen to search for $\Xi$ hyperons, for which the
primary and the secondary decay vertices are clearly separated
(Fig.~\ref{scheme} (bottom)). The $\Xi$ hyperons were identified using a two
dimensional Closest Distance of Approach (CDA) procedure. The CDA values were
calculated between the $\Lambda$($\bar{\Lambda}$)~line of flight and a charged
particle track not associated to the primary vertex. The $\Lambda$ baryons were
taken from the samples shown in Fig.~\ref{inv-mass-nocoll}. A collinearity cut
was then imposed on the direction of the $\Xi$ momentum and the line connecting
it to the primary vertex. The value of the cut, $\theta_{col}<0.02$ rad, is
larger than the value used for $\Sigma^{*}$ reconstruction, since the direction
of the $\Xi$ is reconstructed less precisely than that for $\Lambda$. The
invariant mass distributions are shown in Fig.~\ref{xi_mass_minus_before}. The
resulting resolutions for $\Xi$ and $\bar\Xi$ are the same:
$2.8\pm0.1$~MeV/$c^2$.

In order to extract the yield ratios of heavy hyperons to $\Lambda$ baryons, the
ratios of the corresponding acceptances had to be evaluated. 
It should be noted that the acceptance corrections were evaluated only for the region $Q^{2}>1~$~(GeV/{\it c})$^{2}$. 
The calculation was done using a Monte Carlo simulation based on the LEPTO 6.5.1 generator for
DIS events with default parameters, and a full spectrometer description based on
GEANT 3.21. For each hyperon, the acceptance was calculated as the ratio of
$N_{rec}$, the number of reconstructed hyperons, and $N_{gen}$, the number of
hyperons generated by LEPTO. The same reconstruction and selection procedure
were used as for the real data.

The resulting values of the acceptance ratios for $\Sigma^{*}$ to $\Lambda$ and
for $\Xi$ to $\Lambda$ are 0.67 and 0.42, respectively. The difference between
$\Sigma^{*}$ and $\Xi$ acceptance ratios is explained by different decay
patterns: the $\Sigma^{*}$ hyperons decay practically at the primary vertex,
while for the $\Xi$ hyperons there exists a secondary one. The acceptance ratio
also includes a correction for the branching ratio,
Br($\Sigma^{*}\rightarrow\Lambda\pi$) = 0.88$\pm$0.02~\cite{PDG}. 
It should be noted that Figs.~\ref{inv-mass-nocoll},~\ref{sigma_mass},~\ref{xi_mass_minus_before} are not acceptance-corrected.

The invariant mass distributions for $\Sigma^{*+}$ and $\bar{\Sigma}^{*-}$
(Fig.~\ref{sigma_mass} (top)) were fitted by a sum of a signal function, $S(x)$,
described by a convolution of a Breit-Wigner and a Gaussian, and a background
function $B(x)$:

\begin{equation} \label{frm::peak}
S(x) = \frac{\Gamma}{(2\pi)^{3/2}} \int \frac{N e^{-\frac{1}{2} (\frac{t - x}{\sigma})^2}}{(t-M)^2 + 
(\frac{\Gamma}{2})^2}~dt~;
\end{equation}
\begin{equation} \label{frm::background}
B(x) = a~(x - M_{th})^b~e^{-c~(x - M_{th})^d}.
\end{equation}

The two other invariant mass distributions, $\Lambda\pi^{-}$ and $\bar{\Lambda}\pi^{+}$, 
include contributions coming from the $\Xi$ decay (Fig.~\ref{sigma_mass} (bottom)). 
These contributions were taken into account by adding a second Gaussian function to the signal.
The values of hyperon mass $M$ and width $\Gamma$ were fixed to the PDG values ~\cite{PDG}. 
The value of $M_{th} = 1254$~MeV was chosen to be the sum of $\Lambda$ and $\pi$ masses, and 
$a,~b,~c$ and $d$ were free parameters.

The invariant mass distributions shown for $\Xi^{-}$ and $\bar{\Xi}^{+}$ in 
Fig.~\ref{xi_mass_minus_before} were fitted by a sum of a Gaussian function for 
the signal and a function $B(x)$ for the background, described by an analogous parameterisation as 
the one used for $\Sigma^{*}$, given in Eq.~(\ref{frm::background}). 
It should be noted that a first study of $\Xi$ production using COMPASS 
data was done for the pentaquark $\Phi$(1860) search~\cite{Penta}.

\section{Discussion of results}
\label{sec:discussions} 
The ratios of the acceptance-corrected yields of $\Sigma^{*}$ and $\Xi$ to that
of $\Lambda$ hyperons are given in Table \ref{res} along with their statistical
and systematic uncertainties. Three sources of systematic uncertainties were
considered:\\ a) The uncertainty on the number of hyperon events was estimated
by varying the width of the window for the selection of $\Lambda$ and
$\bar{\Lambda}$ samples, from $\pm 2\,\sigma$ to $\pm 2.5\,\sigma$ and to $\pm
1.5\,\sigma$. This variation results in differences of 0.003, 0.002, 0.002, and
0.002 for the relative yields of $\Sigma^{*+}$, $\bar{\Sigma}^{*-}$,
$\Sigma^{*-}$, and $\bar{\Sigma}^{*+}$, respectively. The corresponding values
for the relative yields of $\Xi^{-}$ and $\bar{\Xi}^{+}$ are equal to 0.001.\\
b) The systematic uncertainties coming from the evaluation of the background
were estimated using a mixed event method. In this method, the shape of the
background in the $\Lambda\pi$ invariant-mass distribution was determined by
combining lambdas and pions from different events. The energies of these pions
were chosen to be similar to the energy of the pion from the $\Lambda$ decay.
The standard collinearity cut ($\theta_{col}$) was also applied. The
uncertainties resulting from this procedure are 0.003, 0.004, 0.004, and 0.005
for the relative yields of $\Sigma^{*+}$, $\bar{\Sigma}^{*-}$, $\Sigma^{*-}$,
and $\bar{\Sigma}^{*+}$, respectively. The uncertainties for the relative yields
of $\Xi^{-}$ and $\bar{\Xi}^{+}$ were found to be negligible. \\ c) The
systematic uncertainties on the acceptance arising from tuning the Monte Carlo
parameters were evaluated to be 0.003, 0.004, 0.005, and 0.003 for
$\Sigma^{*+}$, $\bar{\Sigma}^{*-}$, $\Sigma^{*-}$, and $\bar{\Sigma}^{*+}$,
respectively. For $\Xi^{-}$ and $\bar{\Xi}^{+}$ these uncertainties are
0.002.\\ The combined systematic uncertainties were calculated by summing
quadratically these three contributions.

The experimental ratios show that the number of heavier hyperons compared to
that of $\Lambda$ hyperons is small, in the range 3.8\% to 5.6\%. The results
also indicate that the percentage of $\Lambda$ originating from the decay of
$\Sigma^{*}$ and $\Xi$ hyperons is almost the same (within quoted uncertainties)
as the percentage of $\bar{\Lambda}$ originating from the decay of the
respective antiparticles.

The ratios of production yields of hyperons and antihyperons to those of
$\Lambda$ and $\bar{\Lambda}$ are obtained for the first time in charged lepton
DIS reactions. Earlier, only hyperon to $\Lambda$ yields, but no yields for
antiparticles, were measured in neutrino DIS by the NOMAD
Collaboration~\cite{NOMAD-resonances1}. 
The NOMAD values are also shown in Table~\ref{res}; the average neutrino energy of charged current
interactions was 45.3 GeV. 
We note that COMPASS has collected considerably larger (from
30 to 130 times) samples of hyperons than NOMAD in the current fragmentation
region. It is interesting to compare the charged lepton and the neutrino data
despite the different underlying interactions. COMPASS measures similar values
for the $\Sigma^{*+}/\Lambda$ and $\Sigma^{*-}/\Lambda$ ratios. Taking into
account experimental uncertainties, the same conclusion is valid for NOMAD data
but with NOMAD values being a factor of two smaller than the COMPASS ones.
Finally, within uncertainties, the $\Xi^-/\Lambda$ yield ratio measured by NOMAD
is consistent with zero, while COMPASS gives comparable and non-zero values for
$\Xi^-/\Lambda$ and $\bar{\Xi}^{+}/\bar{\Lambda}$ ratios. The large
experimental uncertainties in the NOMAD measurements prevent us from drawing
conclusions about heavy hyperon production in charged lepton DIS as compared to
neutrino DIS.

\begin{table}[!htb]
\caption{The heavy hyperon to $\Lambda$ and antihyperon to
$\bar{\Lambda}$ yield ratios in lepton DIS. The results are compared with those from 
NOMAD ~\cite{NOMAD-resonances1} neutrino DIS data in the current fragmentation region.$^3$ 
The average neutrino energy of charged current interactions was 45.3 GeV. }
\begin{center}
\begin{tabular}{lcc}
\hline
Ratios & This work & NOMAD \\
\hline
$\Sigma^{*+}/ \Lambda$ & $0.055\pm0.005(stat)\pm0.005(syst)$ & $0.025\pm0.019$\\
$\bar{\Sigma}^{*-}/\bar{\Lambda}$ & $0.047\pm 0.006(stat)\pm0.006(syst)$ & - \\
$\Sigma^{*-} / \Lambda $ & $ 0.056\pm 0.009(stat)\pm0.007(syst)$ & $0.037\pm0.015$\\
$\bar{\Sigma}^{*+}/\bar{\Lambda}$ & $ 0.039\pm 0.006(stat)\pm0.006(syst)$ & - \\
$\Xi^- /\Lambda$ & $0.038 \pm 0.003(stat) \pm 0.002(syst)$ & $0.007\pm0.007$\\
$\bar{\Xi}^+ / \bar{\Lambda}$ & $0.043\pm 0.004(stat)\pm 0.002(syst)$ & -\\
\hline
 \end{tabular}
 \label{res}
\end{center} 
\end{table}
\footnotetext[3]{The quoted numbers of NOMAD are not corrected for acceptance.
Nevertheless, as shown in Ref.~\cite{NOMAD-resonances1}, the acceptance
uncorrected ratios $\Sigma^{*+}/\Lambda$ and $\Sigma^{*-}/\Lambda$ in the full
$x_F$ region are practically the same as the corrected ones. As a good
approximation one may expect the same behavior for the current fragmentation
region.}

In a different approach, the same COMPASS yield ratios as discussed above were
also evaluated after removing the DIS cuts $Q^2 > 1$~(GeV/$c$)$^2$ and $0.2 < y
<0.9$. Only the initial selection for $\Lambda$($\bar{\Lambda}$) candidates was
applied: a) events with two oppositely charged hadron tracks form the secondary
vertex, b) hadrons with momenta larger than 1~GeV/$c$, c) $p_T>23$ MeV/$c$ on
the transverse momentum of the decay products with respect to the hyperon
direction, d) $\Lambda$($\bar{\Lambda}$) candidates in the current fragmentation
region $x_F>0.05$.

The resulting $\Lambda$($\bar{\Lambda}$) samples are about ten times larger than
those obtained when using DIS cuts. In total, $N(\rm \Lambda \rightarrow
p\pi^-) = 1208413\pm1312$ and $N(\rm \bar \Lambda \rightarrow p\pi^+) =
654387\pm1067$ events were reconstructed. The $\Sigma^{*}$ hyperon signals are
also enhanced. 
The number of $\Sigma^{*}$ resulting 
from the fits are: $N(\Sigma^{*+})=44780 \pm 1301$, $N(\Sigma^{*-})=22716 \pm 872$, 
$N(\bar{\Sigma}^{*-})=37728 \pm 1361$ and $N(\bar{\Sigma}^{*+})=19813\pm 1169$. 
The numbers of $\Xi$ hyperons are: $N(\Xi^{-}) = 20458 \pm 162$ and 
$N(\bar{\Xi}^{+}) = 11448 \pm 128$. 
The invariant mass distributions for $\Lambda$, $\Sigma^{*+}$, $\Sigma^{*-}$, 
$\Xi^{-}$ and their antiparticles 
without DIS cuts are given in~\ref{sec:appl}.
In Table~\ref{com} the relative heavy hyperons yield ratios 
obtained using DIS and non-DIS $\Lambda$($\bar{\Lambda}$) 
samples are given. One can see that within the experimental 
uncertainties the  yield ratios for both samples are compartible. 
\begin{table}[!htb]
\caption{Heavy hyperon to $\Lambda$ and antihyperon to $\bar{\Lambda}$ yield ratios
without DIS cuts normalized to the same ratios with DIS cuts.}
\begin{center}
 \begin{tabular}{lc}
\hline
& Relative yield ratios without$/$with DIS cuts \\
\hline
$\Sigma^{*+} / \Lambda$  & $ 1.03\pm 0.08(stat)$ \\
$\bar{\Sigma}^{*-}/\bar{\Lambda}$ & $ 0.97\pm 0.11(stat)$ \\
$\Sigma^{*-} / \Lambda  $ & $ 1.03\pm 0.16(stat)$ \\
$\bar{\Sigma}^{*+}/\bar{\Lambda}$ & $ 0.97\pm 0.13(stat)$ \\
$\Xi^-/\Lambda$  & $1.06\pm 0.09(stat)$ \\
$\bar{\Xi}^+ / \bar{\Lambda}$ & $1.06\pm 0.09(stat)$ \\
\hline
 \end{tabular}
 \end{center}
\label{com}
\end{table}

The average $Q^2$ for this sample drops to $\langle Q^2 \rangle $ = 0.47 $(\mbox{GeV}/c)^2$, as
compared to $\langle Q^2 \rangle = 3.58~(\mbox{GeV}/c)^2$ when using the DIS
cut. This observation indicates that the measured yield ratios are not strongly
depending on $Q^2$. A check of the $y$ dependence of the results was also made.
The $y$ interval was divided in 2 bins, larger and smaller than $y=0.5$. The
ratios were calculated in these bins with and without $Q^2$ cut. In each $y$
bin the ratios with and without $Q^2$ cut are compatible within statistical
uncertainties. The ratios in the large-$y$ bin show a tendency to be on average
$\sim 15\%$ higher than those in the small-$y$ bin.

\begin{table}[!htb]
\caption{The heavy hyperon to $\Lambda$ yield ratios in DIS.}
\begin{center}
\renewcommand{\arraystretch}{1.2}
 \begin{tabular}{llll}
\hline
 Ratios   & LEPTO  & COMPASS data & LEPTO \\
     & (Default) &     & (COMPASS) \\
 \hline
$\Lambda/\bar{\Lambda}$ & $ 1.22\pm0.01$& $ 1.71\pm0.02$ & $1.72\pm0.01$\\
$K^{0} / \Lambda$ & $ 6.06\pm0.01$& $ 6.21\pm0.05$ & $6.22\pm0.01$\\
\hline
$\Sigma^{*+}/\Lambda $ & $ 0.082\pm0.001$& $ 0.055\pm0.005$& $0.052\pm0.001$\\
$\bar{\Sigma}^{*-} / \bar{\Lambda} $ & $ 0.074\pm0.001$ & $ 0.047 \pm 0.006$ & $0.038\pm0.001$\\
$\Sigma^{*-} / \Lambda $ & $ 0.084\pm0.001 $ & $ 0.056\pm 0.009$ & $0.067\pm0.001$\\
$\bar{\Sigma}^{*+} / \bar{\Lambda} $ & $ 0.060\pm0.001$ & $ 0.039\pm 0.006$ & $0.037\pm0.001$ \\
\hline
$\Xi^- / \Lambda $ & $ 0.051\pm0.001 $ & $ 0.038\pm 0.003$ & $0.029\pm0.001$\\
$\bar{\Xi}^+ / \bar{\Lambda} $ & $ 0.056\pm0.001$ & $ 0.043\pm 0.004$ & $0.040\pm0.001$\\
\hline
$\Sigma^{0}/ \Lambda $ & $ 0.200\pm0.003 $ & $-$ & $0.130\pm0.002$\\
$\bar{\Sigma}^{0}/ \bar{\Lambda} $ & $ 0.200\pm0.003 $ & $-$ & $0.120\pm0.002$\\\hline
\end{tabular}
\end{center}
\label{rcorr2}
\end{table}

The ratios of production yields between heavy hyperons and $\Lambda$ particles
are important for the interpretation of the results on the longitudinal
polarisation transfer in DIS. Indeed, a $\Lambda$ hyperon originating from the
decay of a heavier hyperon is polarised differently than the directly produced
$\Lambda$ particle. The indirectly produced $\Lambda$ mainly come from the
decay of $\Sigma^{0}$, $\Sigma^{*+}$, and $\Xi$ hyperons. In
Ref.~\cite{HERMES-lambda}, the contribution of the indirectly produced $\Lambda$
was estimated by a Monte Carlo simulation to be as large as 60$\%$. Our Monte
Carlo simulation with LEPTO default parameters shows that this contribution is
about 58$\%$ for $\Lambda$ and 54$\%$ for $\bar{\Lambda}$. With tuned LEPTO
parameters (discussed further below) the fractions of the indirectly produced
$\Lambda$ and $\bar{\Lambda}$ are reduced to $37\%$ and $32\%$,
respectively.

Only the contributions from charged heavy hyperons were considered in the
present analysis. The contribution from radiative decay $\Sigma^{0} \rightarrow
\Lambda + \gamma$ can only be indirectly estimated using the LEPTO simulation
code, in which the final-state hadronisation is described by the Lund string
fragmentation model. The production yield ratios calculated with the LEPTO
default parameters are given in the first column of Table~\ref{rcorr2}. Their
comparison with the COMPASS results given in the second column shows that this
simulation overestimates the experimental ratios for heavy hyperons by about
$\sim 1.5$. The $\Lambda$ to $\bar \Lambda$ ratio exhibits an opposite trend
whereas the $K^{0}/\Lambda$ ratio is close to the experimental value.

In order to reproduce better the measured ratios given in Table~\ref{rcorr2},
the LEPTO/ JETSET~7.4 parameters~\cite{JETSET} related to the production yields
of strange baryons were tuned (see Table~\ref{FF}). These parameters
characterize the properties of the LEPTO generator not associated with kinematic
distributions of hyperons: PARJ(1) - suppression of diquark-antidiquark pair
production in the colour field; PARJ(2) - suppression of $s \bar{s}$-pair
production compared to $u \bar{u}$- or $d \bar{d}$-pair production; PARJ(3) -
extra suppression of strange diquark production compared to the normal
suppression of strange quarks; PARJ(4) - suppression of spin-1 diquarks compared
to spin-0 ones; PARJ(5) - relative occurrence of baryon-antibaryon production;
PARJ(7) - strange meson suppression factor. 
A study of the MC distributions of the common SIDIS $Q^2$ and $W$ 
and the baryon variables $z$ and $p_T$ for 
$\Lambda$, $\Sigma^{*}$, $\Xi$ and their antiparticles was performed. The
distributions of two MC data sets, with default and tuned parameters, were found
to be consistent within errors. The $Q^2$, $W$, $z$ and $p_T$ ratios of real
data and Monte Carlo samples with both tuned and default parameters are similar
without strong deviations from unity. 

The simulated results obtained with the tuned parameters are shown in the third
column of Table~\ref{rcorr2}. The measured ratios of the heavy hyperon to
$\Lambda$ yields are now well reproduced. In addition, the agreement between
the data and LEPTO for the $\Lambda$ to $\bar \Lambda$, and $K$ to $\Lambda$
ratios is now very good. Finally, the new parameters also modify the unmeasured
$\Sigma^0/\Lambda$ ratio.
 
For completeness, the acceptance corrections were recalculated using the newly
tuned LEPTO parameters. The new and old corrections agree within one standard
deviation. The difference was included in the systematic uncertainties,
mentioned at the beginning of this Section.

\begin{table}[!htb]
\caption{The default and COMPASS-tuned LEPTO/JETSET parameters.}
 \begin{center}
  \begin{tabular}{lll}
  \hline
  Parameters & Default & COMPASS\\
  \hline
  PARJ(1) & 0.1 & 0.03 \\ 
  PARJ(2) & 0.3 & 0.45 \\ 
  PARJ(3) & 0.4 & 0.175 \\  
  PARJ(4) & 0.05 & 0.078 \\
  PARJ(5) & 0.5 & 3.0 \\ 
  PARJ(7) & 0.5 & 0.13 \\ 
  \hline
  \end{tabular}
  \end{center}
  \label{FF}
 \end{table}


\section{\label{sec:concl} Conclusions}
The heavy hyperon to $\Lambda$ and heavy antihyperon to $\bar{\Lambda}$ yield
ratios were measured for the first time in charged lepton deep-inelastic
scattering. All yield ratios were found to be in the range 3.8\% to 5.6\%.
Within the relative uncertainties of about 10\%, the yield ratios for hyperons
and antihyperons are quite similar. No strong $Q^2$ dependence of the ratios
was found within the statistical accuracy. The obtained results imply that some
parameters of the LEPTO code, which are associated with strange quark production
and fragmentation in charged lepton DIS processes, should be substantially
modified. Using the tuned LEPTO parameters, the fractions of indirectly
produced $\Lambda$ and $\bar{\Lambda}$ hyperons were found to be $37\%$
and $32\%$, respectively.

\section*{Acknowledgements}
We gratefully acknowledge the support of the CERN management and staff, the
skill and effort of the technicians of our collaborating institutes. Special
thanks are due to V. Anosov and V. Pesaro for their technical support during
installation and running of this experiment. It is a pleasure to thank
S. Belostotsky, D. Naumov and Yu. Naryshkin for stimulating discussions.

\appendix
\section{\label{sec:appl} The sample without DIS cuts}
\begin{figure}[h!]
\begin{center}
\begin{tabular}{c}
\includegraphics[width=0.44\textwidth]{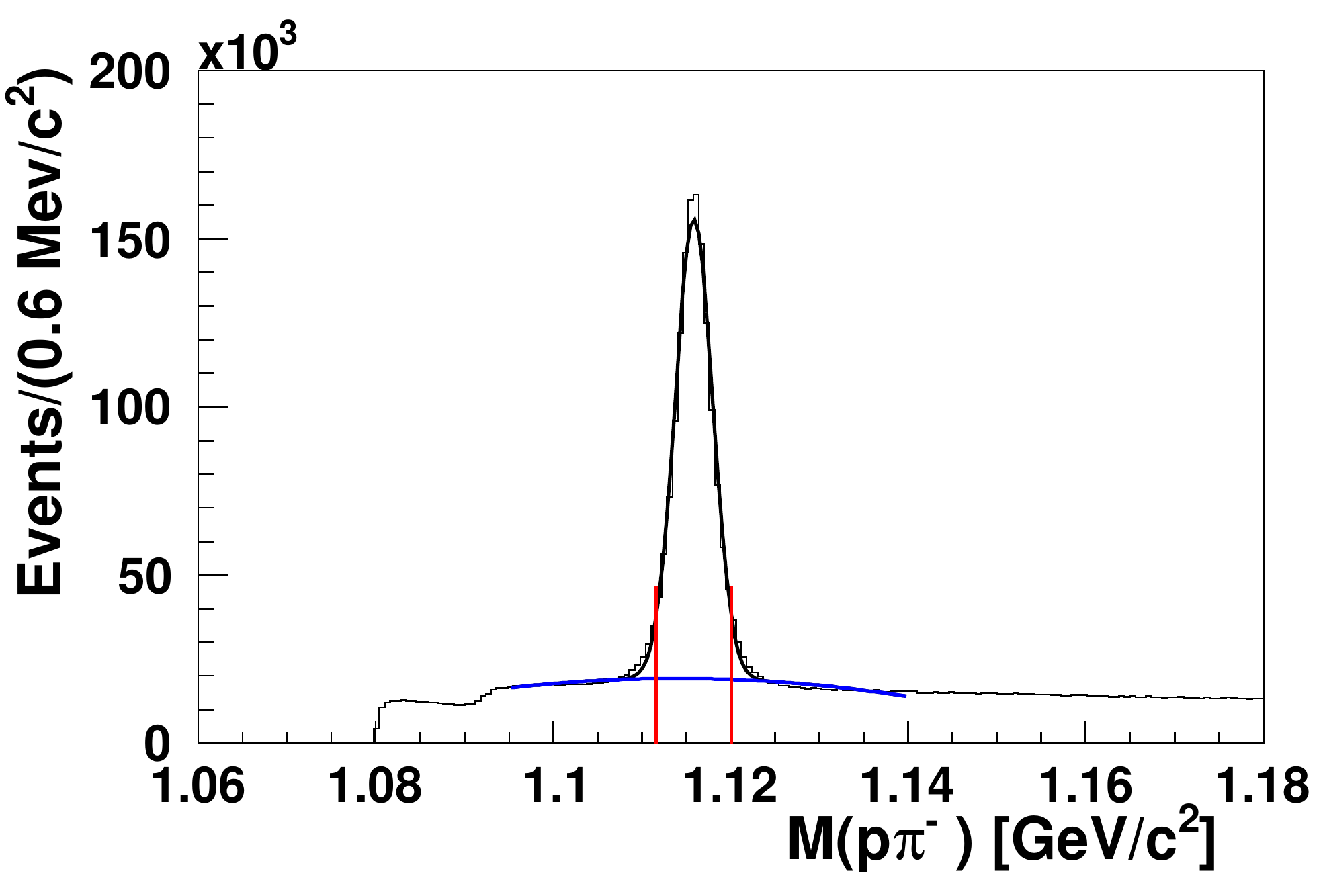} \\
\includegraphics[width=0.44\textwidth]{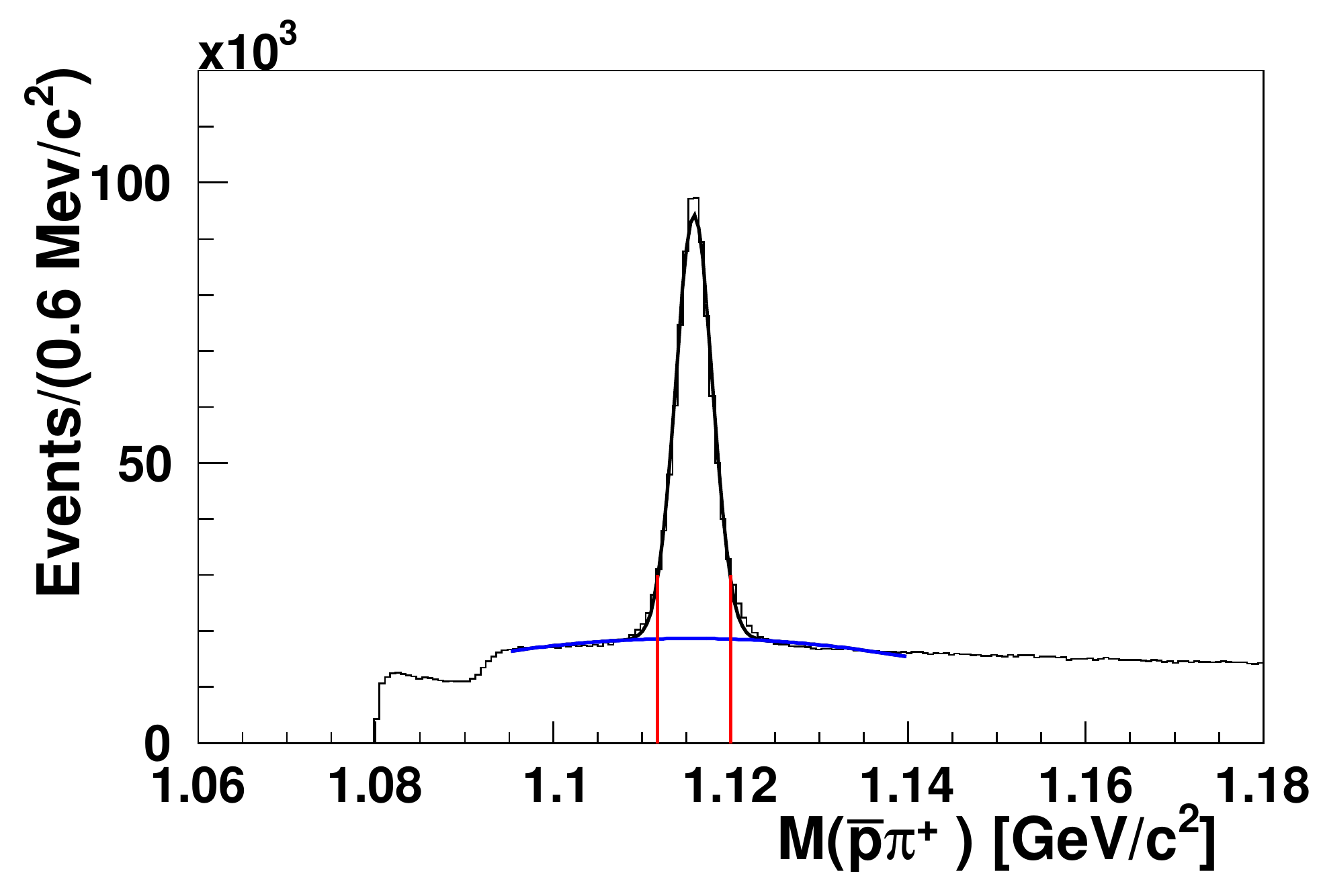} 
\end{tabular}
\caption{The $p\pi^-$ (top) and $\bar{p}\pi^+$ (bottom) invariant mass
distributions without DIS cuts. The total numbers of $\Lambda$ and $\bar{\Lambda}$ determined
within the fit interval are $N(\Lambda)=1208413\pm1312$ and $N(\bar \Lambda)=654387\pm1067$.}
\label{inv-mass-nocoll-noDIS}
\end{center}
\end{figure}

\begin{figure*}[h!]
\begin{center}
\begin{tabular}{cc}
\includegraphics[width=0.44\textwidth]{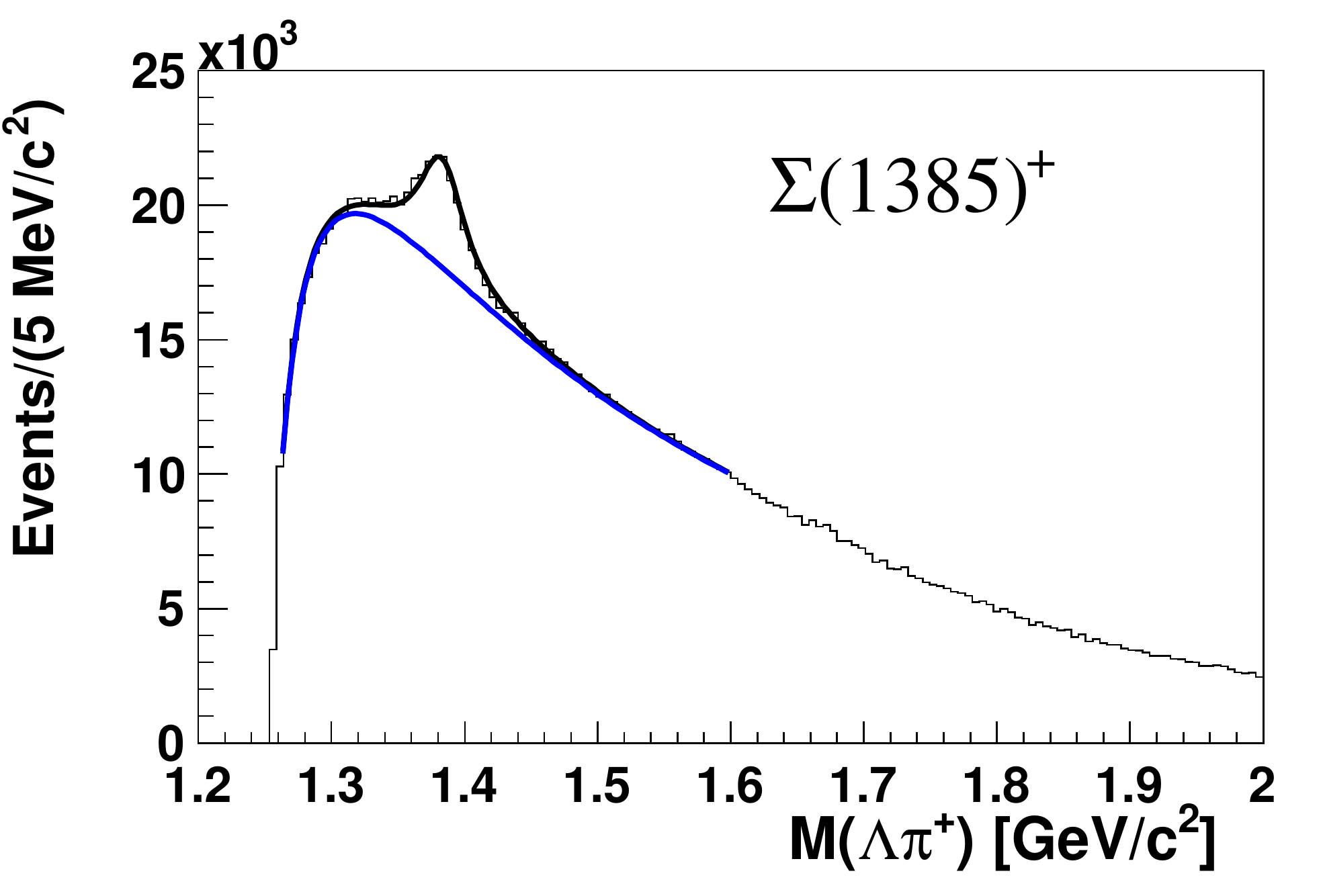} &
\includegraphics[width=0.44\textwidth]{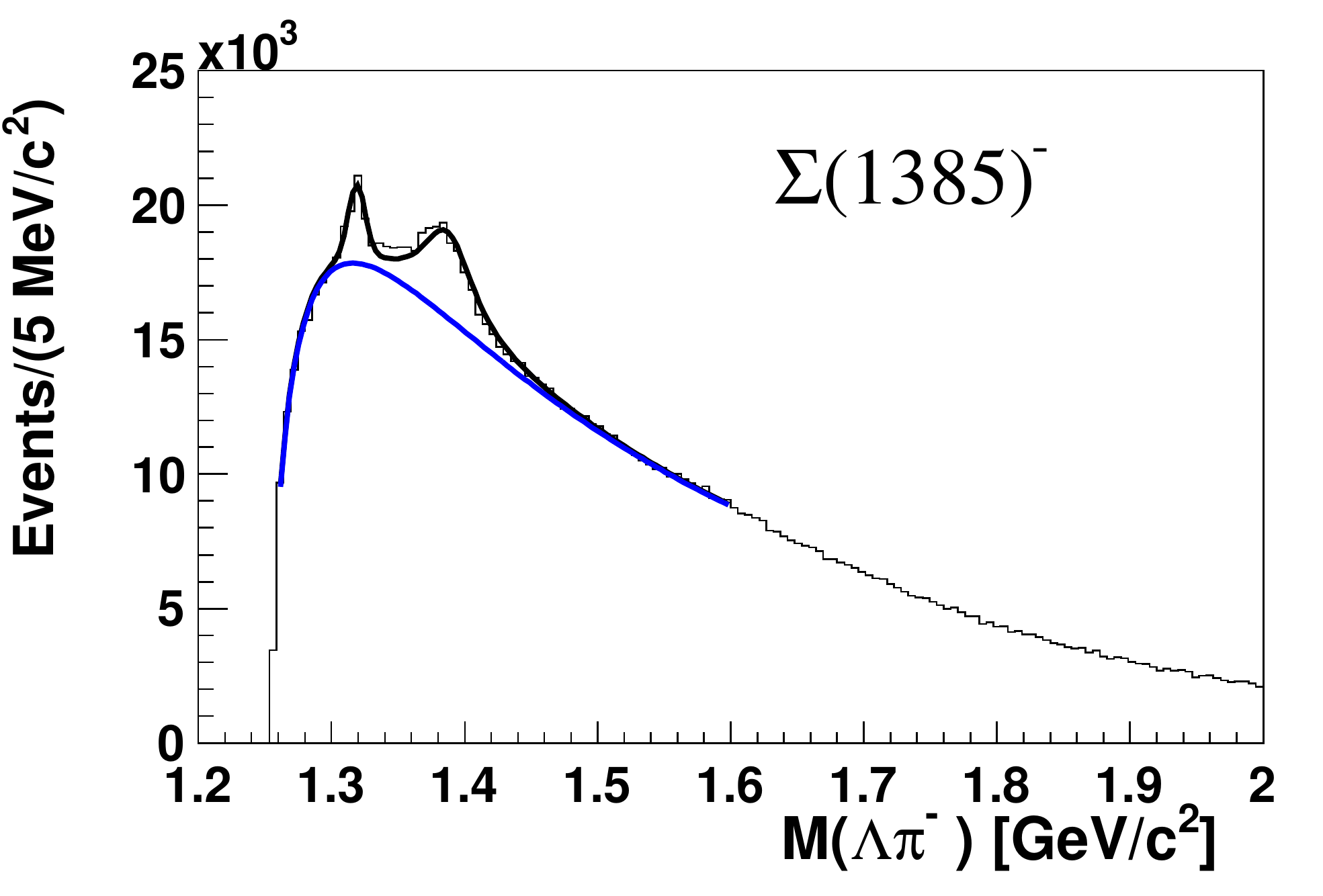} \\
a) & b) \\
\includegraphics[width=0.44\textwidth]{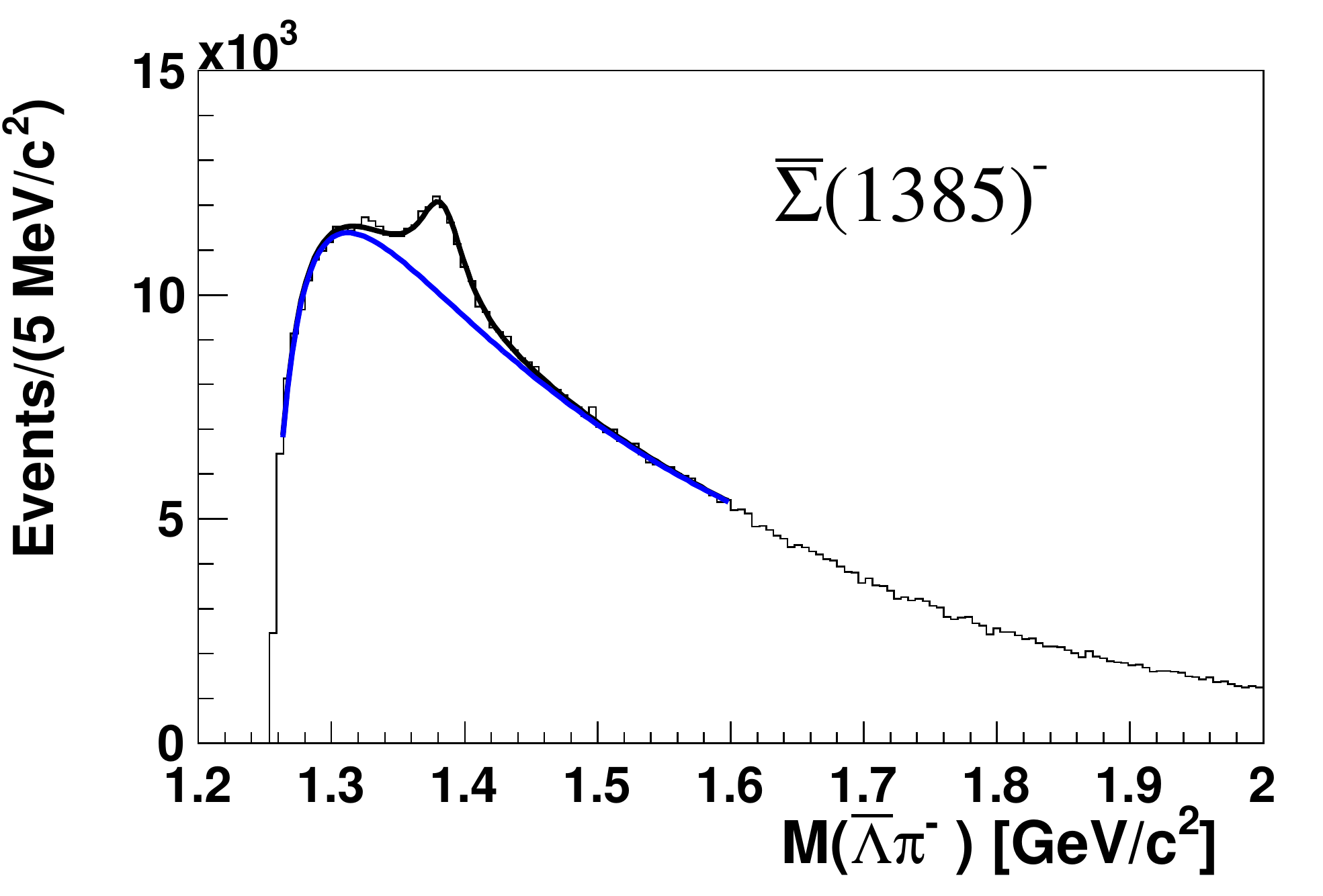} &
\includegraphics[width=0.44\textwidth]{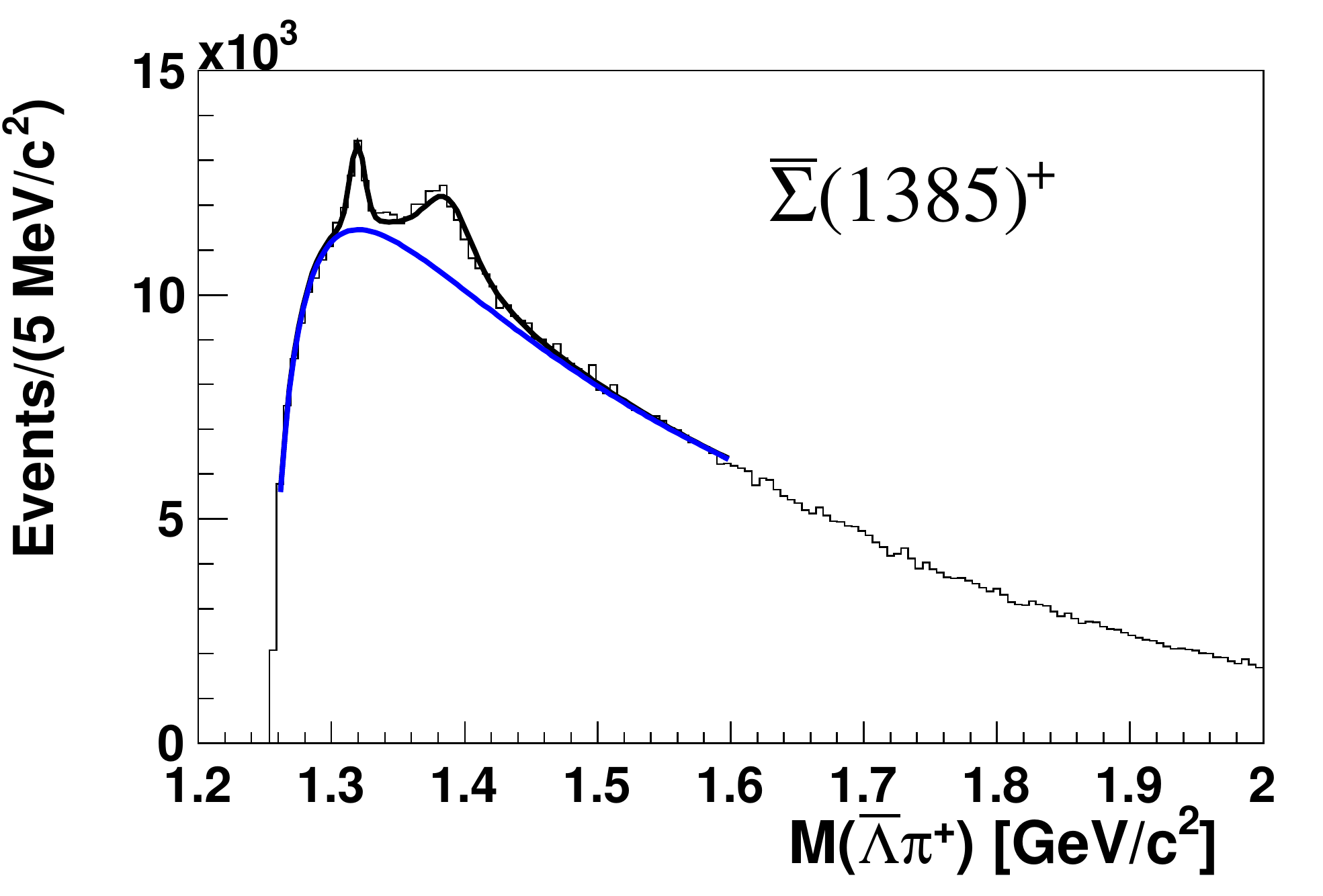} \\
c) & d) \\
\end{tabular}
\caption{The $\Lambda\pi$ invariant mass distributions without DIS cuts. The solid lines
represent the signal plus backgroud and the background only obtained from the
fit. The number of $\Sigma^{*}$ resulting
from the fits are: a) $N(\Sigma^{*+})=44780 \pm 1301$, b) $N(\Sigma^{*-})=22716 \pm 872$,
c) $N(\bar{\Sigma}^{*-})=37728 \pm 1361$ and d) $N(\bar{\Sigma}^{*+})=19813\pm 1169$.}
\label{sigma_mass_noDIS}
\end{center}
\end{figure*}

\begin{figure*}[h!]
\begin{center}
\begin{tabular}{cc}
\includegraphics[width=0.44\textwidth]{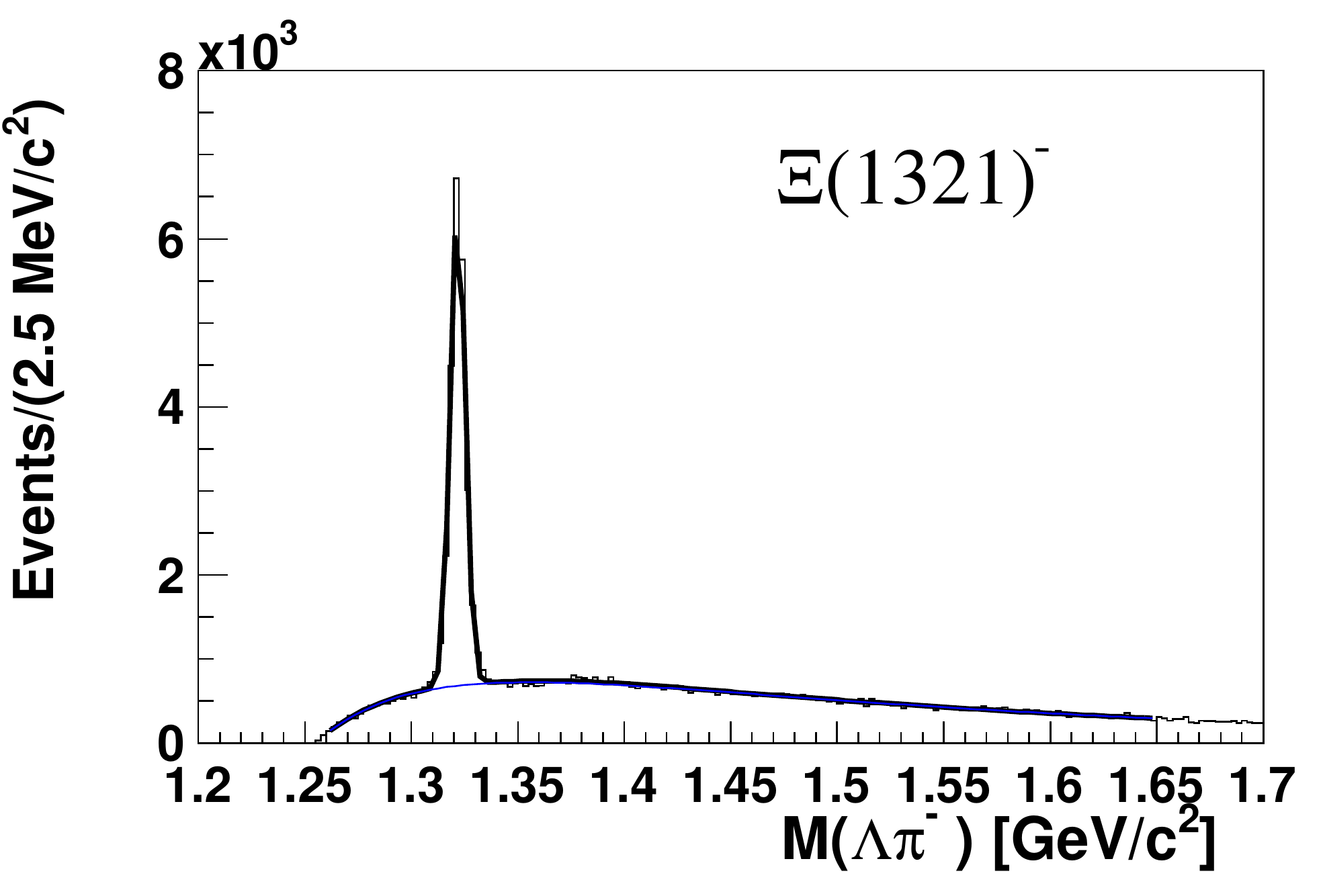}&
\includegraphics[width=0.44\textwidth]{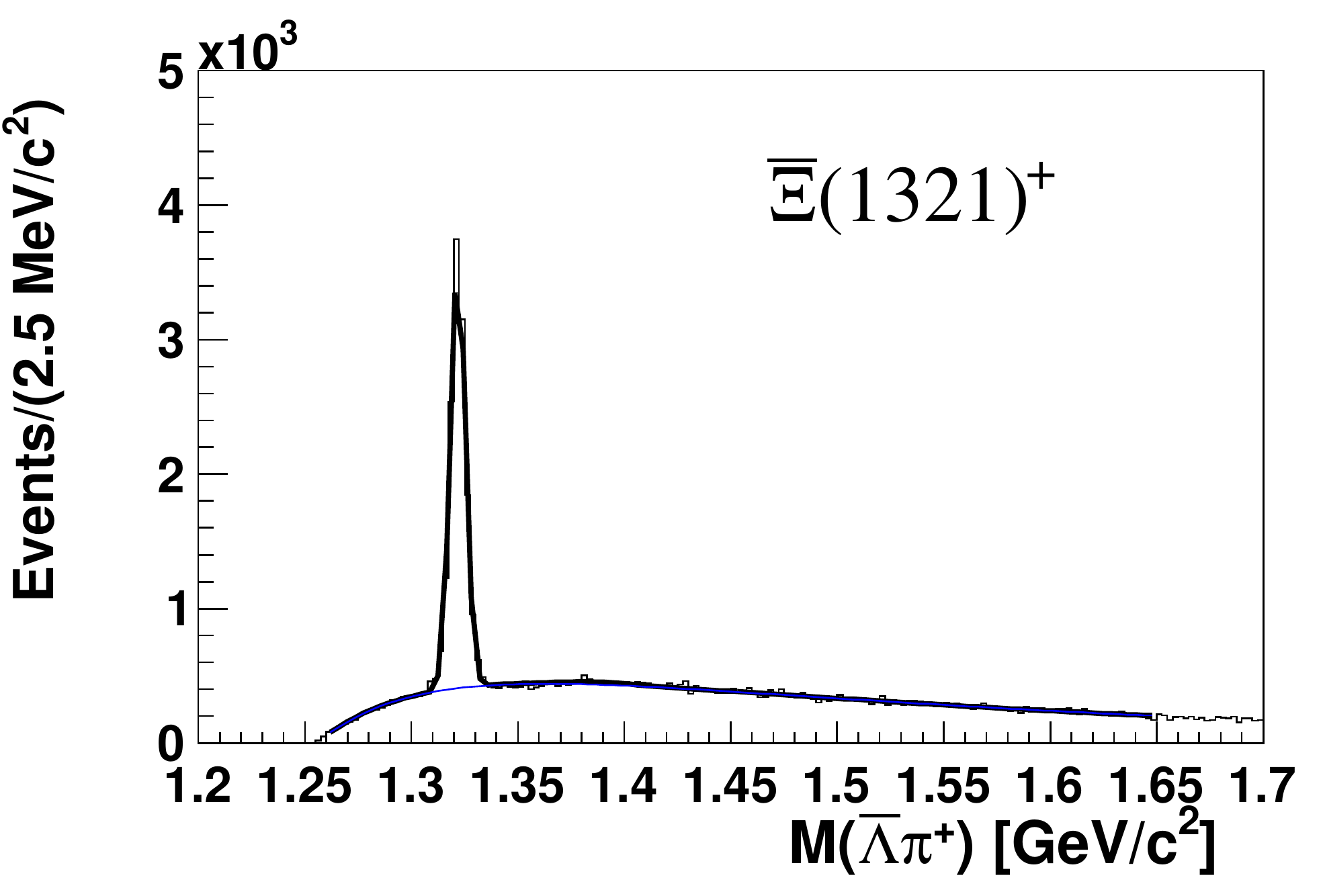} \\
\end{tabular}
\caption{Invariant mass distributions for $\Lambda\pi^{-}$ (left) and
$\bar{\Lambda}\pi^{+}$ (right) pairs without the DIS cuts. The solid lines represent the signal plus
backgroud and the background only obtained from the fit. The estimated numbers of 
$\Xi$ hyperons are: $N(\Xi^{-}) = 20458 \pm 162$ and 
$N(\bar{\Xi}^{+}) = 11448 \pm 128$.}
\label{xi_mass_minus_before_noDIS}
\end{center}
\end{figure*}

The  cuts on the four-momentum squared of the virtual photon, $Q^2 > 1$~(GeV/$c$)$^2$,
and on the fractional energy $y$ of the virtual photon, $0.2 < y < 0.9$ 
were ommited here but all other cuts  were kept as in the DIS sample. 
The invariant mass distributions for $\Lambda$, $\Sigma^{*+}$, $\Sigma^{*-}$,$\Xi^{-}$ and their antiparticles 
are shown in Figs.~\ref{inv-mass-nocoll-noDIS},~\ref{sigma_mass_noDIS},~\ref{xi_mass_minus_before_noDIS}.

\begin{thebibliography}{99}
\bibitem{HERMES-lambda} A.~Airapetian et al. (HERMES Collaboration),
  \emph{Phys. Rev.} {\bf D74} 072004 (2006)
\bibitem{Compass:09} M.~Alekseev et al. (COMPASS Collaboration),
  \emph{Eur. Phys. J.} \textbf{64} 171 (2009)
\bibitem{LEPTO} A.E.G.~Ingelman, A.~Edin, J.~Rathsman,
  \emph{Comp. Phys. Commun.} \textbf{101}  108 (1997)
\bibitem{NOMAD-resonances1} P.~Astier et al. (NOMAD Collaboration),
  \emph{Nucl.Phys.} {\bf B621} 3 (2002)
\bibitem{setup} P.~Abbon et al. (COMPASS Collaboration),
  \emph{Nucl. Instrum. Meth.} \textbf{A577} 455 (2007)
\bibitem{NOMAD-resonances2} P.~Astier et al. (NOMAD Collaboration),
  \emph{Nucl. Phys.} {\bf B588} 3 (2000) 
\bibitem{NOMAD-resonances3} P.~Astier et al. (NOMAD Collaboration),
  \emph{Nucl. Phys.} {\bf B605} 3 (2001)
\bibitem{E665} M.R.~Adams et al. (E665 Collaboration), \emph{Eur. Phys. J.} {\bf
  C17}  263 (2000)
\bibitem{STAR} Quinhua Xu (STAR Collaboration), Proc.17th International 
Spin Physics Symposium (SPIN06), Kyoto, AIP conference Proceedings, 915 (2006) 428, hep-ex/0612035
\bibitem{PDG} Review of Particle Properties, \emph{J. Phys.} \textbf{G 37}, 7A
  075021 (2010)
\bibitem{Penta} E.S.~Ageev et al. (COMPASS Collaboration), \emph{Eur. Phys. J.}
  \textbf{C 41},  469 (2005)
\bibitem{JETSET} {T.~Sj\"ostrand}, \emph{Comp. Phys. Commun.} \textbf{82}  74 (1994)
\end{thebibliography}
\end{document}